\documentclass[12pt]{article}
\usepackage{graphics,epsfig,here,color}

\definecolor{Blue}{named}{Blue}
\definecolor{Red}{named}{Red}
\definecolor{Green}{named}{ForestGreen}
\definecolor{Black}{named}{Black}
\definecolor{Olive}{named}{OliveGreen}
\definecolor{Royal}{named}{RoyalBlue}
\definecolor{Orange}{named}{YellowOrange}
\definecolor{Yellow}{named}{Goldenrod}
\definecolor{Cornblue}{named}{CornflowerBlue}
\definecolor{Lila}{named}{DarkOrchid}

\def\beq   {\begin{equation}}
\def\eeq   {\end{equation}}
\def\beqd  {\begin{displaymath}}
\def\eeqd  {\end{displaymath}}
\def\beqaa {\begin{eqnarray}}
\def\eeqaa {\end{eqnarray}}

\def\ti  {\tilde}

\def\sz{\ifmmode{\tilde{\chi}^0} \else{$\tilde{\chi}^0$} \fi}
\def\sw{\ifmmode{\tilde{\chi}} \else{$\tilde{\chi}$} \fi}

\newcommand{\be}[1]{\begin{equation} \label{(#1)}}
\newcommand{\ee}{\end{equation}}
\newcommand{\baq}[1]{\begin{eqnarray} \label{(#1)}}
\newcommand{\eaq}{\end{eqnarray}}
\newcommand{\rf}[1]{(\ref{(#1)})}
\newcommand{\ba}{\begin{array}}
\newcommand{\ea}{\end{array}}

\newcommand{\slashed}[1]{\not\!#1}

\begin{document}
\pagestyle{empty}

\vspace*{-1cm} 
\begin{flushright}
CERN-PH-TH/2005-137\\ 
DCPT-05-84\\
IPPP-05-42\\
TSL/ISV-2005-0295\\
UWThPh-2005-14 \\
WUE-ITP-2005-011\\
\end{flushright}

\vspace*{1.4cm}

\begin{center}

{\Large
{\bf CP-odd observables in neutralino production 
with transverse $e^+$ and $e^-$ beam polarization
}}

\vspace{2cm}

{\large 
A.~Bartl$^{1}$, H.~Fraas$^{2}$, S.~Hesselbach$^{3}$, K.~Hohenwarter-Sodek$^{1}$, 
T.~Kernreiter$^{1}$, G.~Moortgat-Pick$^{4}$}
\end{center}
\vspace{1cm}

{\it \noindent $^{1}$~Institut f\"ur Theoretische Physik, Universit\"at Wien, A-1090
Vienna, Austria\\
$^{2}$~Institut f\"ur Theoretische Physik und Astrophysik, Universit\"at W\"urzburg,\\ 
\phantom{$^{2}$ }D-97074 W\"urzburg, Germany\\
$^{3}$~High Energy Physics, Uppsala University, Box 535, S-75121 Uppsala, Sweden\\
$^{4}$~TH Division, Physics Department, CERN, CH-1211 Geneva 23, Switzerland}


\vspace{1cm}

\begin{abstract} 
We consider neutralino production and decay $e^+e^-\to\ti\chi^0_i\ti\chi^0_j$, 
$\tilde{\chi}^0_j\to \tilde{\chi}^0_1 f \bar{f}$
at a linear collider with transverse $e^+$ and $e^-$ beam polarization.
We propose CP asymmetries by means of the azimuthal
distribution of the produced neutralinos and 
of that of the final leptons,
while taking also into account the subsequent 
decays of the neutralinos. We include the
complete spin correlations between production and decay.
Our framework is the Minimal Supersymmetric 
Standard Model with complex parameters.
In a numerical study we show that there are
good prospects to observe these CP asymmetries
at the International Linear Collider and estimate the accuracy
expected for the determination of the phases in the 
neutralino sector.
\end{abstract}

\newpage
\pagestyle{plain}


\section{Introduction}

Supersymmetry (SUSY) is at present one of the most prominent
extensions of the Standard Model (SM) \cite{Haber:1984rc}.
It can solve the hierachy problem, allows the unification of the gauge couplings 
and has the additional merit of providing new sources of CP violation.
In the chargino and neutralino sectors of the Minimal Supersymmetric Standard 
Model (MSSM), the higgsino mass parameter $\mu$ and 
the gaugino mass parameter $M_1$ are in general complex,
while the $SU(2)$ gaugino mass parameter 
$M_2$ can be chosen real by redefining the fields.
The precise determination of the underlying SUSY
parameters will be one of the main goals of the high-luminosity 
$e^+e^-$ International Linear Collider (ILC)~\cite{TDR}.

The phases of the complex parameters $\mu$ and $M_1$ may be
constrained or correlated by the experimental upper bounds on
the electric dipole moments (EDM) of electron, neutron and the atoms
$^{199}$Hg and $^{205}$Tl \cite{edmexp}. 
These constraints, however, are rather model
dependent. In a constrained MSSM the restrictions on the phases
of $\mu$ and $M_1$ can be rather severe. However, there may be
cancellations between the different SUSY contributions to the EDMs,
which allow larger values for the phases (for reviews see, e.g. 
\cite{edmstheo}). 
If $\mu$ and $M_1$ are
complex and all other parameters are real, in general the phase
of $\mu$ has to be small. However, the restrictions on the phase
of $\mu$ may disappear if also lepton flavour violating terms
in the MSSM lagrangian are included \cite{Bartl:2003ju}. 
It is, therefore, necessary
to determine in an independent way the phases of 
the complex SUSY parameters by measurements
of suitable CP-sensitive observables.   
Experiments with transverse $e^{\pm}$
beam polarization may allow us to construct suitable observables for precision
studies of the effects of new physics and CP violation.
Recent studies on the advantages of transversely polarized 
$e^+$ and $e^-$ beams  
are presented in \cite{Budny:tk,Ananthanarayan:2003wi,Power}.

The study of neutralino production 
\begin{equation}
\label{eq:prodneut}
e^+e^-\to\ti\chi^0_i\ti\chi^0_j ~,\qquad i,j=1,\dots,4~,
\end{equation}
and subsequent two-body decay processes
\begin{eqnarray}
&&\tilde{\chi}^0_j \to \ell^{\pm} \ti \ell^{\mp}_n \to 
\ell^{\pm} \ell^{\mp} \ti\chi^0_1~,\qquad n=1,2~,
\label{eq:decaychain} 
\end{eqnarray}
with $\ell=e,\mu$, will play an important role at the ILC. 
The production process has
been studied extensively  in the literature 
(see \cite{TDR} and references therein). 
Production and subsequent
decay processes, and the decay angular and energy
distributions have been studied in detail in 
\cite{Bartl:1986hp}--\cite{Moortgat-Pick:1999di}. 
The properties of Majorana and Dirac particles and their
production and decay amplitudes under CP and CPT have been studied in
\cite{Petcov:1984nf}--\cite{Choi:2005gt}.

In \cite{Choi:1998ei}--\cite{Choi:2001ww} 
it has been shown that the parameters
of the chargino and neutralino systems can be determined
by measuring suitable CP-even observables. However,
the measurement of CP-odd observables is necessary to clearly
demonstrate that CP is violated and for
the unambiguous determination of the CP-violating phases.
Several T-odd observables in neutralino production and
decay applying triple product 
correlations have been proposed in \cite{Choi:1999cc,Bartl:2003tr}.

Transversely polarized beams offer the possibility
to construct further CP-sensitive observables. 
This is the subject of the present paper. 
We propose and study CP-odd asymmetries for the case of
transverse $e^+$ and $e^-$ beam polarizations.
We define two types of CP asymmetries, one that involves
only neutralino production, the other one involving
production and decay if one of the neutralinos decays as in 
Eq.~(\ref{eq:decaychain}). 
Note that in the case of chargino
production it has been shown that the analogous CP-odd asymmetries
vanish \cite{Bartl:2004xy}.  In neutralino production it is possible,
however, to construct CP-odd asymmetries involving the
transverse beam polarization, because there are $t$-channel {\it
and} $u$-channel contributions due to the Majorana nature of the
neutralinos \cite{Ananthanarayan:2003wi}. 
The formulae for the production cross section of process (\ref{eq:prodneut}),
for longitudinally and transversely polarized beams, have
been given in \cite{Choi:2001ww,Chiappetta:1985nb}. 
In this paper we derive the compact analytic
formulae for the CP asymmetries 
with the help of the spin density matrix formalism 
\cite{spinhaber}, including the complete spin 
correlations between production and decay 
of the neutralinos. We study numerically the parameter and
phase dependences of the CP asymmetries.

The paper is organized as follows: in Section \ref{lagrangian} we set up 
the definitions. We present the formulae for the cross section
of (\ref{eq:prodneut}) with transverse beam polarization 
in Section \ref{crossection}.
In Section \ref{cpasymmetries} we define the CP-odd asymmetries. 
We present a numerical investigation of these asymmetries
in Section \ref{numerics}, while
Section \ref{conclusion} contains our conclusions.

\section{Lagrangian and couplings \label{lagrangian}}

\begin{figure}[H]
\hspace{-2.5cm}
\begin{minipage}[t]{6cm}
\begin{center}
{\setlength{\unitlength}{0.6cm}
\begin{picture}(5,5)
\put(-2.5,-8.5){\includegraphics{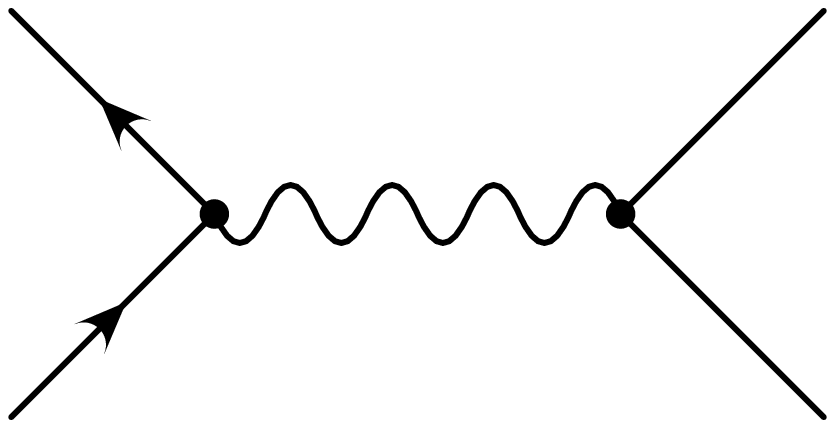}}
\put(1.2,-.4){{\small $e^{-}(p_1)$}}
\put(7.8,-.4){{\small $\tilde{\chi}^0_j(p_4)$}}
\put(1.2,3.8){{\small $e^{+}(p_2)$}}
\put(7.8,3.8){{\small $\tilde{\chi}^0_i(p_3)$}}
\put(5.1,2.5){{\small $Z^0$}}
\end{picture}}
\end{center}
\end{minipage}
\hspace{-0.5cm}
\begin{minipage}[t]{5cm}
\begin{center}
{\setlength{\unitlength}{0.6cm}
\begin{picture}(2.5,5)
\put(-4,-9){\includegraphics{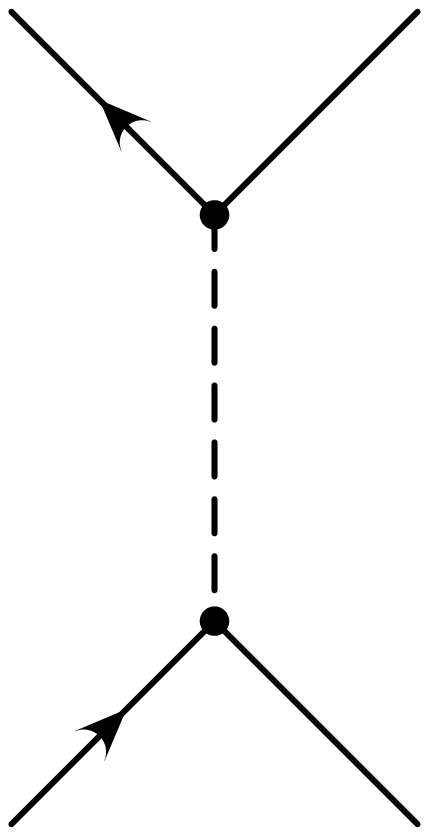}}
\put(.8,-1.5){{\small $e^{-}(p_1)$}}
\put(.8,4.8){{\small $e^{+}(p_2)$}}
\put(5.8,-1.5){{\small $\tilde{\chi}^0_j(p_4)$}}
\put(5.8,4.8){{\small $\tilde{\chi}^0_i(p_3)$}}
\put(4.4,1.5){{\small $\tilde{e}_{L,R}$}}
 \end{picture}}
\end{center}
\end{minipage}
\begin{minipage}[t]{5cm}
\begin{center}
{\setlength{\unitlength}{0.6cm}
\begin{picture}(2.5,5)
\put(-4.5,-9){\includegraphics{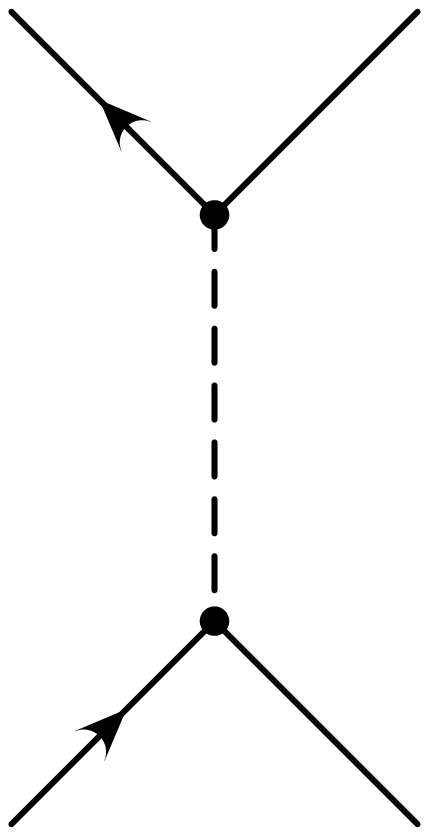}}
\put(.3,-1.5){{\small $e^{-}(p_1)$}}
\put(.3,4.8){{\small $e^{+}(p_2)$}}
\put(5.5,-1.5){{\small $\tilde{\chi}^0_i (p_3)$}}
\put(5.5,4.8){{\small $\tilde{\chi}^0_j (p_4)$}}
\put(3.9,1.5){{\small $\tilde{e}_{L,R}$}}
\end{picture}}
\end{center}
\end{minipage}
\vspace{.7cm}
\caption{\label{Fig:FeynProd}Feynman diagrams of the production process
  $e^{+}e^{-}\to\tilde{\chi}^0_i\tilde{\chi}^0_j$.}
\end{figure}

The tree-level Feynman diagrams for the production process (\ref{eq:prodneut})
are given in Fig.~\ref{Fig:FeynProd}. 
$\vspace{0.25cm}$
$\hspace{-0.75cm}$ The interaction Lagrangians are \cite{Haber:1984rc}
\begin{equation}
\mathcal{L}_{Z^0\ell^+\ell^-}=-\frac{g}{\cos\Theta_W}Z_{\mu}\bar{\ell}
\gamma^{\mu}[L_\ell P_L+R_\ell P_R]\ell~,
\end{equation}
\begin{equation}
\mathcal{L}_{Z^0\tilde{\chi_i}\tilde{\chi_j}}=\frac{1}{2}
\frac{g}{\cos\Theta_W}Z_{\mu}{\bar{\tilde{\chi}}}^0_i
\gamma^{\mu}[O''^L_{ij}P_L+O''^R_{ij}P_R]{\tilde{\chi}}^0_j~,
\end{equation}
\begin{equation}
\mathcal{L}_{\ell\tilde{\ell}\tilde{\chi}_i}=
gf^L_{\ell i}\bar{\ell}P_R\tilde{\chi}^0_i\tilde{\ell}_L
+gf^R_{\ell i}\bar{\ell}P_L\tilde{\chi}^0_i\tilde{\ell}_R+{\rm h.c.} 
\end{equation}
with $i,j=1,\dots,4\,$ and the couplings
\begin{equation}
f^L_{\ell i}=-\sqrt{2}\left[\frac{1}{\cos\Theta_W}\left(-\frac{1}{2}
+\sin^2\Theta_W \right)N_{i2}-\sin\Theta_W N_{i1}\right]~,
\end{equation}
\begin{equation}
f^R_{\ell i}=\sqrt{2}\sin\Theta_W[\tan\Theta_WN^*_{i2}-N^*_{i1}]~,
\end{equation}
\begin{equation}
O''^L_{ij}=-\frac{1}{2}(N_{i3}N^*_{j3}-N_{i4}N^*_{j4})\cos2\beta
-\frac{1}{2}(N_{i3}N^*_{j4}+N_{i4}N^*_{j3})\sin2\beta~,
\end{equation}
\begin{equation}
O''^R_{ij}=-O''^{L*}_{ij}~,
\end{equation}
\begin{equation}
L_\ell=-\frac{1}{2}+\sin^2\Theta_W,\quad R_\ell=\sin^2\Theta_W ,
\end{equation}
where $P_{L,R}=\frac{1}{2}(1\mp\gamma_5)$.
$g$ is the weak coupling constant
($g=e/\cos\Theta_W$, $e>0$), $\Theta_W$ is the
weak mixing angle and $\tan\beta=v_2/v_1$
is the ratio of the
vacuum expectation values of the Higgs fields.
The unitary $(4\times4)$ matrix $N_{ij}$
diagonalizes the complex symmetric 
neutralino mass matrix in the basis 
($\ti\gamma,\ti Z, \ti H^0_a,\ti H^0_b$) \cite{Bartl:1986hp}.

\section{Cross section \label{crossection}}

\begin{figure}[t]
\setlength{\unitlength}{1cm}
\begin{picture}(12,6)
\put(1.5,0.0)
{\mbox{\epsfysize=6cm\epsffile{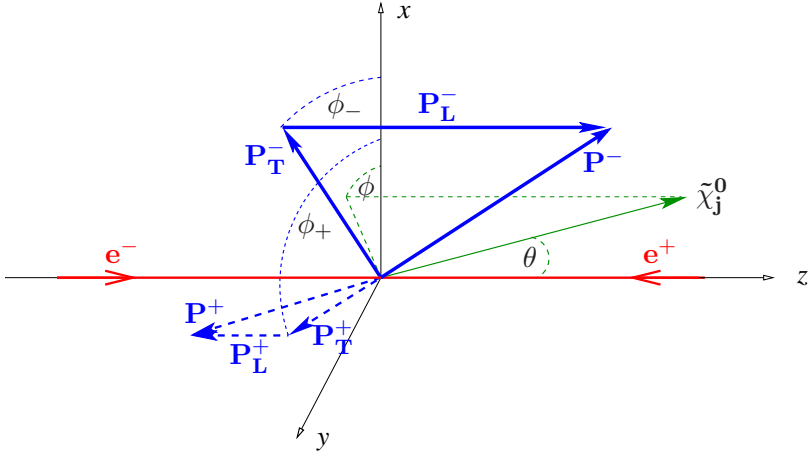}}}
\put(10.7,3.3){\small{${\bf \ti{\chi}^0_j}$}}
\put(10.,2.5){\small\color{red}{${\bf e^+}$}}
\put(8.4,2.45){\small{${\bf \theta}$}}
\put(5.4,2.9){\small{${\bf \phi_+}$}}
\put(5.8,4.5){\small{${\bf \phi_-}$}}
\put(6.2,3.4){\small{${\bf \phi}$}}
\put(2.85,2.5){\small\color{red}{${\bf e^-}$}}
\put(4.7,3.8){\small\color{blue}{${\bf P^-_T}$}}
\put(7,4.5){\small\color{blue}{${\bf P^-_L}$}}
\put(9.2,3.7){\small\color{blue}{${\bf P^-}$}}
\put(5.6,1.4){\small\color{blue}{${\bf P^+_T}$}}
\put(4.5,1.15){\small\color{blue}{${\bf P^+_L}$}}
\put(3.9,1.7){\small\color{blue}{${\bf P^+}$}}
\end{picture}
\caption{Decomposition of the $e^{\pm}$ polarization vectors ${\bf P^\pm}$
into the longitudinal components ${\bf P^\pm_L}$ 
in the direction of the electron/positron momentum  and the 
transverse components ${\bf P^\pm_T}={\cal P}^\pm_T \vec{t}_\pm$
with respect to a fixed coordinate system $(x,y,z)$. The
$z$-axis is in the direction of the electron momentum.  
\label{fig_skizze}}
\end{figure}

In order to calculate the squared amplitude of  
process (\ref{eq:prodneut}) with subsequent decay (\ref{eq:decaychain}), 
we use the spin density matrix formalism \cite{spinhaber}.
The squared amplitude of the combined process of production and decay reads
\begin{equation}
|T|^2=\sum_{\lambda_i\lambda_j\lambda_i^{'}\lambda_j^{'}}
|\Delta(\tilde{\chi}^0_i)|^2 |\Delta(\tilde{\chi}^0_j)|^2
\rho_P^{\lambda_i \lambda_j, \lambda^{'}_i \lambda^{'}_j}
\rho_{D, \lambda^{'}_i\lambda_i}
\rho_{D,\lambda_j^{'} \lambda_j}~,
\label{eq4_4e}
\end{equation}
where $\Delta(\tilde{\chi}^0_{i,j})$ is the propagator 
of the corresponding neutralino;
$\rho_P^{\lambda_i \lambda_j, \lambda^{'}_i \lambda^{'}_j}$ denotes the spin 
density matrix of the production, 
$\rho_{D, \lambda^{'}_{i,j}\lambda_{i,j}}$ 
are the spin density matrices of the decay; $\lambda_{i,j}$ denotes 
the helicity of the neutralino $\ti{\chi}^0_{i,j}$. 
The propagators are given by
\begin{equation}
\Delta(\tilde{\chi}^0_{k})=1/[p^2_{\tilde{\chi}_k}-m_{\tilde{\chi}_k}^2+i m_{\tilde{\chi}_k} 
\Gamma_{\tilde{\chi}_k}]~, \;\,\, k=i,j~.
\end{equation}
Here $p_{\tilde{\chi}_k}$, $m_{\tilde{\chi}_k}$ and $\Gamma_{\tilde{\chi}_k}$ denote the
four-momentum, mass and total width of the neutralino
$\tilde{\chi}^0_{k}$. For these propagators we use the narrow-width
approximation.
The (unnormalized) spin density production matrix is given by
\begin{equation}
\rho_P^{\lambda_i\lambda_j, \lambda_i^{'}\lambda_j^{'}}=
\sum_{\lambda_{e^-} \lambda_{e^+} \lambda_{e^-}^{'} \lambda_{e^+}^{'}}
\rho(e^-)_{\lambda'_{e^-}\lambda_{e^-}}\rho(e^+)_{\lambda'_{e^+} \lambda_{e^+}}
T_{P, \lambda_{e^-} \lambda_{e^+}}^{\lambda_i\lambda_j}
T_{P, \lambda'_{e^-} \lambda'_{e^+}}^{\lambda_i^{'}\lambda_j^{'}*}~, 
\label{eq_rev1}
\end{equation}
where $T_{P, \lambda_{e^-} \lambda_{e^+}}^{\lambda_i\lambda_j}$ is the
helicity amplitude of the production process and
$\lambda_{e^{\pm}}$ is the helicity of $e^\pm$.
The spin density decay matrices can be written as
\begin{eqnarray}
\rho_{D, \lambda_i^{'} \lambda_i} & = & T_{D, \lambda_i}T_{D,\lambda'_i}^{*}~,
\\
\rho_{D, \lambda_j^{'} \lambda_j} & = & T_{D, \lambda_j}T_{D,\lambda'_j}^{*}~,
\label{eq_rev2}
\end{eqnarray}
where $T_{D, \lambda_{i,j}}$ is the helicity amplitude for the 
decay.
The spin density matrices of the polarized 
$e^+$ and $e^-$ beam in Eq.~(\ref{eq_rev1}) can be written as
\begin{equation}
\rho(e^{\pm}) =  \frac{1}{2}(1+{\cal P}^\pm_i\sigma^i)~, 
\label{eq_eldensity}
\end{equation}
where ${\cal P}^\pm_1$ is the degree of
transverse $e^{\pm}$ beam polarization in the production
plane, ${\cal P}^\pm_2$ is the degree of transverse $e^\pm$ beam 
polarization perpendicular
to the production plane, 
${\cal P}^\pm_3={\cal P}^\pm_L \,
(-1 \leq {\cal P}^\pm_L \leq 1)$ is the degree of longitudinal
$e^\pm$ beam polarization and $\sigma^i$ ($i=1,2,3$) are the Pauli matrices.
For the degrees of transverse beam polarizations we have the relation    
$({\cal P}^{\pm}_1)^2+({\cal P}^{\pm}_2)^2=({\cal P}^{\pm}_T)^2$ 
with ${\cal P}^{\pm}_1=\cos\phi_{\pm} {\cal P}^{\pm}_T$
and ${\cal P}^{\pm}_2=\sin\phi_{\pm} {\cal P}^{\pm}_T$ 
($0\leq {\cal P}^{\pm}_T \leq 1$ and  
$({\cal P}^\pm_L)^2+({\cal P}^{\pm}_T)^2 \leq 1$), 
see Fig. \ref{fig_skizze}
\footnote[1]{At this point we note that,
contrary to the usual conditions at a circular accelerator, where the
Sokolov--Ternov effect orientates automatically both transverse
polarization vectors either parallel or antiparallel (depending on the
sign of the charge of the incoming particle), there is the
possibility, at the ILC, to choose an {\it arbitrary} transverse
polarization for both $e^+$ and $e^-$, independent from each other.}.

The production and decay matrices are calculated with the help of
the Bouchiat--Michel formulae~\cite{bouchiat}:
\begin{eqnarray}
u(p,\lambda^{'})\bar{u}(p,\lambda)&=&
\frac{1}{2}[\delta_{\lambda \lambda^{'}} + \gamma_5 {\slashed s}^a
\sigma^a_{\lambda \lambda^{'}}](\slashed p+m)~,\label{eq_bouch1}\\
v(p,\lambda^{'})\bar{v}(p,\lambda)&=&
\frac{1}{2}[\delta_{\lambda^{'} \lambda} + \gamma_5 {\slashed s}^a
\sigma^a_{\lambda^{'}\lambda}](\slashed p-m)~.
\label{eq_bouch2}
\end{eqnarray}
The
three four-component spin basis vectors $s^a$ and the 4-vector $p/m$ form an
orthonormal system.  
For the incoming particles $e^+$ and $e^-$ in the limit of
vanishing electron mass, $m_e\to 0$,
Eqs.~(\ref{eq_bouch1}) and (\ref{eq_bouch2}) can be written as
\begin{equation}
\mbox{\hspace{-.5cm}}
\hspace{0.5cm} u(p_{e^-}, \lambda_{e^-})
\bar{u}(p_{e^-},\lambda^{'}_{e^{-}})= \frac{1}{2}\{(1+2
\lambda_{e^-} \gamma_5) \delta_{\lambda^{'}_{e^-}
\lambda_{e^-}}+ \gamma_5 [\slashed
t^{\,\,1}_{e^-}\sigma^{1}_{\lambda^{'}_{e^-} \lambda_{e^-}}
+\slashed t^{\,\,2}_{e^-}\sigma^2_{\lambda^{'}_{e^-}\lambda_{e^-}}
]\}\slashed p_{e^-},
\label{eq_bouch1-he}
\end{equation}
\begin{equation} 
\mbox{\hspace{-.5cm}}
\hspace{0.5cm} v(p_{e^+}, \lambda_{e^+}^{'})
\bar{v}(p_{e^+},\lambda_{e^+})= \frac{1}{2}\{ (1-2
\lambda_{e^+} \gamma_5) 
\delta_{\lambda^{'}_{e^+}\lambda_{e^+}} + \gamma_5 [\slashed
t^{\,\,1}_{e^+}\sigma^1_{\lambda^{'}_{e^+} \lambda_{e^+}}+
\slashed t^{\,\,2}_{e^+}
\sigma^2_{\lambda^{'}_{e^+}\lambda_{e^+}}]\}\slashed p_{e^+},
\label{eq_bouch2-he}
\end{equation}
where $t_{e^\pm}^1$ and $t_{e^\pm}^2$ 
are the basis 4-vectors of the transverse
polarization of the electron and positron beam, respectively.
Thus, the transverse polarization 4-vectors can be written as 
\begin{equation}
t_{{\pm}}=\cos(\phi_{\pm}-\phi)t_{e^{\pm}}^1+
\sin(\phi_{\pm}-\phi) t_{e^{\pm}}^2~,
\label{def-trans-basis}
\end{equation}
where $\phi$ is the azimuthal angle of the scattering plane and 
$\phi_{\pm}$ are the azimuthal angles of the transverse polarization
with respect to a fixed
reference system, see Fig.~\ref{fig_skizze}.
A convenient choice of $t_{e^{\pm}}^1$ and $t_{e^{\pm}}^2$ 
in the laboratory system is given 
in Appendix~\ref{appendixA}.

With Eqs.~(\ref{eq_bouch1})--(\ref{eq_bouch2-he}), 
the spin density production matrix and the spin density decay matrices can be 
expanded in terms of the Pauli matrices $\sigma^a$ and $\sigma^b$,
where the superscripts $a$ $(b)=1,2,3$ refer to the
polarization vectors of $\tilde{\chi}^0_i$ ($\tilde{\chi}^0_j$):
\begin{eqnarray}
\rho_P^{\lambda_i\lambda_j, \lambda_i^{'}\lambda_j^{'}}&=&
\delta_{\lambda_i\lambda_i^{'}} \delta_{\lambda_j\lambda_j^{'}}
P(\tilde{\chi}^0_i\tilde{\chi}^0_j)
+\delta_{\lambda_j\lambda_j^{'}}\sum_{a=1}^3
\sigma^a_{\lambda_i\lambda_i^{'}}\Sigma^a_P(\tilde{\chi}^0_i)
\nonumber\\ &&
+\delta_{\lambda_i\lambda_i^{'}}\sum_{b=1}^3
\sigma^b_{\lambda_j\lambda_j^{'}}\Sigma^b_P(\tilde{\chi}^0_j)
+\sum_{a,b=1}^3\sigma^a_{\lambda_i\lambda_i^{'}}
\sigma^b_{\lambda_j\lambda_j^{'}}
\Sigma^{ab}_P(\tilde{\chi}^0_i\tilde{\chi}^0_j)~,\label{eq4_4h}\\
\rho_{D,\lambda_i^{'}\lambda_i}&=&\delta_{\lambda_i^{'}\lambda_i}
D(\ti{\chi}^0_i)+\sum_{a=1}^3
\sigma^a_{\lambda_i^{'}\lambda_i} \Sigma^a_D(\ti{\chi}^0_i)~,\label{eq4_4i}\\
\rho_{D,\lambda_j^{'}\lambda_j}&=&\delta_{\lambda_j^{'}\lambda_j}
D(\ti{\chi}^0_j)+\sum_{b=1}^3
\sigma^b_{\lambda_j^{'}\lambda_j} \Sigma^b_D(\ti{\chi}^0_j)~.\label{eq4_4j}
\end{eqnarray}
The contribution $P(\tilde{\chi}^0_i\tilde{\chi}^0_j)$ is independent
of the neutralino polarization, whereas
$\Sigma^a_P(\tilde{\chi}^0_i)$ and
$\Sigma^b_P(\tilde{\chi}^0_j)$ depend on the polarization
of the corresponding neutralino.
Then
$\Sigma^3_P(\tilde{\chi}^0_{i,j})/P(\tilde{\chi}^0_i\tilde{\chi}^0_j)$
gives the longitudinal polarization of the neutralino $\ti{\chi}^0_{i,j}$;
$\Sigma^1_P(\tilde{\chi}^0_{i,j})/P(\tilde{\chi}^0_i\tilde{\chi}^0_j)$ is
the transverse polarization of the neutralino in the scattering plane; and
$\Sigma^2_P(\tilde{\chi}^0_{i,j})/P(\tilde{\chi}^0_i\tilde{\chi}^0_j)$
is the polarization perpendicular to the scattering plane.
The terms $\Sigma^{ab}_P(\tilde{\chi}^0_i\tilde{\chi}^0_j )$ describe
the spin correlations between
the polarizations of the two produced particles.
The contribution of the decay matrix, which is independent of
the neutralino polarization, is denoted by $D(\ti{\chi}^0_{i,j})$, and
$\Sigma^{a,b}_D(\tilde{\chi}^0_{i,j})$ denotes the contribution that depends 
on the neutralino polarization.

With Eqs.~(\ref{eq4_4h})--(\ref{eq4_4j}) the squared amplitude  $|T|^2$,
Eq.~(\ref{eq4_4e}), of the combined process of production and decay, 
for arbitrarily polarized beams, is given by
\begin{eqnarray}
|T|^2&=&4|\Delta(\ti{\chi}^0_i)|^2|\Delta(\ti{\chi}^0_j)|^2\Big[P(\ti{\chi}^0_i \ti{\chi}^0_j) D(\ti{\chi}^0_i) D(\ti{\chi}^0_j)
    +\sum^3_{a=1}\Sigma_P^a(\ti{\chi}^0_i) \Sigma_D^a(\ti{\chi}^0_i) D(\ti{\chi}^0_j)
\nonumber\\&&
+\sum^3_{b=1}\Sigma_P^b(\ti{\chi}^0_j) \Sigma_D^b(\ti{\chi}^0_j)
D(\ti{\chi}^0_i)
    +\sum^3_{a,b=1}\Sigma_P^{ab}(\ti{\chi}^0_i \ti{\chi}^0_j)
 \Sigma^a_D(\ti{\chi}^0_i) \Sigma^b_D(\ti{\chi}^0_j)\Big]~,
\label{eq4_5}
\end{eqnarray}
where the contributions
$P(\ti{\chi}^0_i\ti{\chi}^0_j)$, $\Sigma^a_P(\ti{\chi}^0_i)$, 
$\Sigma^b_P(\ti{\chi}^0_j)$, $\Sigma^{ab}_P(\ti{\chi}^0_i\ti{\chi}^0_j)$ 
of the production spin density matrix
include the terms for arbitrary beam polarizations. 
The differential cross section of the production and decay process is given by
\begin{equation}
{\rm d}\sigma=\frac{1}{2 s}|T|^2 
{\rm dLips}(s,p_i)\label{eq_13}~,
\end{equation}
where ${\rm dLips}(s,p_i)=(2\pi)^4 \delta^4(p_1+p_2-\sum_{i} p_i)
\prod_{i}\frac{{\rm d}^3p_i}{(2\pi)^32E_i}~$; 
see Appendix~\ref{appendixB} for more details.

\subsection{
Contributions to the production spin density matrix 
independent of neutralino polarization}
In the following we regard the contribution $P(\ti{\chi}^0_i\ti{\chi}^0_j)$ 
to the spin density production matrix, see Eq.~(\ref{eq4_4h}),
that is independent of the polarization of
the neutralinos $\ti{\chi}^0_i$ and $\ti{\chi}^0_j$.
We only list the contribution $P(\ti{\chi}^0_i\ti{\chi}^0_j)_T$,
which depends on the transverse $e^\pm$ beam polarization:
%
\begin{equation}
P(\ti{\chi}^0_i\ti{\chi}^0_j)_T=
P(ZZ)_T+P(Z \tilde{e}_L)_T+P(Z \tilde{e}_R)_T+P(\tilde{e}_L \tilde{e}_R)_T,
\label{eq:defprodsum}
\end{equation}
\be{eq:PZZ}
P(ZZ)_{T}=-{\cal P}_T^- {\cal P}_T^{+}~\frac{g^4}{\cos^4\Theta_W}|\Delta(Z)|^2 
L_e R_e (|O^L_{ij}|^2+|O^R_{ij}|^2)r_1~,
\ee
\baq{eq:PZeL}
P(Z \tilde{e}_L)_{T}&=& -{\cal P}_T^- {\cal P}_T^{+}~\frac{1}{2}
\frac{g^2}{\cos^2\Theta_W}R_e \nonumber\\
&&{}\times\{{\rm Re}(
\Delta(Z)f^{L*}_{\ell i}f^L_{\ell j}O^L_{ij}
[\Delta(\tilde{e}_L,t)^*+\Delta(\tilde{e}_L,u)^*])~r_1
\nonumber\\
&&{}+{\rm Im}(
\Delta(Z)f^{L*}_{\ell i}f^L_{\ell j}O^L_{ij}
[\Delta(\tilde{e}_L,t)^*-\Delta(\tilde{e}_L,u)^*])~r_2\}~,
\eaq
\baq{eq:PZeR}
P(Z \tilde{e}_R)_{T}&=& -{\cal P}_T^- {\cal P}_T^{+}~\frac{1}{2}
\frac{g^2}{\cos^2\Theta_W}L_e \nonumber\\
&&{}\times\{{\rm Re}(
\Delta(Z)f^{R*}_{\ell i}f^R_{\ell j}O^R_{ij}
[\Delta(\tilde{e}_R,t)^*+\Delta(\tilde{e}_R,u)^*])~r_1
\nonumber\\
&&{}-{\rm Im}(
\Delta(Z)f^{R*}_{\ell i}f^R_{\ell j}O^R_{ij}
[\Delta(\tilde{e}_R,t)^*-\Delta(\tilde{e}_R,u)^*])~r_2\}~,
\eaq
\baq{eq:PeLeR}
P(\tilde{e}_L \tilde{e}_R)_{T}&=& {\cal P}_T^{-} {\cal P}_T^{+}~\frac{1}{4}
g^4 \nonumber\\
&&{}\!\!\!\!\!\!\!\!\!\!\!\!\!
\times \{
{\rm Re}([
\Delta(\tilde{e}_L,t)\Delta(\tilde{e}_R,u)^*+
\Delta(\tilde{e}_L,u)\Delta(\tilde{e}_R,t)^*]
f^{L*}_{\ell i}f^L_{\ell j}f^{R*}_{\ell i}f^R_{\ell j})~r_1
\nonumber\\
&&{}\!\!\!\!\!\!\!\!\!\!\!\!\!+
{\rm Im}([
\Delta(\tilde{e}_L,t)\Delta(\tilde{e}_R,u)^*-
\Delta(\tilde{e}_L,u)\Delta(\tilde{e}_R,t)^*]
f^{L*}_{\ell i}f^L_{\ell j}f^{R*}_{\ell i}f^R_{\ell j})~r_2\}~,
\eaq 
where we have introduced the notation
\baq{eq:r1}
r_1 &=&[(t_- p_4)(t_+ p_3)+(t_- p_3)(t_+ p_4)]
(p_1 p_2)\nonumber\\
&& +[(p_1 p_4)(p_2 p_3)+(p_1 p_3)(p_2 p_4)-(p_1 p_2)(p_3 p_4)]
(t_- t_+)~,
\eaq
\baq{eq:r2}
r_2&=& \varepsilon^{\mu\nu\rho\sigma}[
t_{+,\mu} p_{1\nu} p_{2\rho} p_{4\sigma}
(t_{-} p_3)+
t_{-,\mu} p_{1\nu}p_{2\rho} p_{3\sigma}(t_+ p_4)\nonumber\\[3mm]
&& +
t_{-,\mu} t_{+,\nu} p_{2\rho} p_{4\sigma}(p_1 p_3)+
t_{-,\mu} t_{+,\nu} p_{1\rho} p_{3\sigma}(p_2 p_4)]~,
\eaq
where
$\varepsilon^{0123}=1$ and  
$\Delta(Z)=
i/(s-m^2_Z)$, 
$\Delta(\tilde{e}_{L,R},t)=
i/(t-m^2_{\tilde{e}_{L,R}})$,
$\Delta(\tilde{e}_{L,R},u)=
i/(u-m^2_{\tilde{e}_{L,R}})$ with $s=(p_1+p_2)^2$, 
$t=(p_1-p_4)^2$, $u=(p_1-p_3)^2$.
The masses of the selectrons and the $Z$-boson are given by
$m_{\tilde{e}_{L,R}}$ and $m_Z$. 
Since we study the process far beyond the $Z$-threshold, 
the $Z$-width can be neglected, and all propagators can be taken as purely
imaginary.

Note that only terms bilinearly dependent on 
transverse beam polarizations appear for $m_e\to 0$, 
because the couplings to $e^+ e^-$ 
are vector- or axial-vector-like \cite{Dass:1975mj,Renard}
(for the $\ti e_{L,R}$ exchange the coupling to $e^+ e^-$
can be brought to that form via Fierz identities \cite{Fierz}). 
Inspecting Eqs.~\rf{eq:PZZ}--\rf{eq:PeLeR}, we note that  
transverse beam polarization gives rise to the interference term 
$P(\ti e_L \ti e_R)_{T}$, which
is absent for longitudinal beam polarization~\cite{Moortgat-Pick:1999di}.
On the other hand, there are no terms $P(\ti e_L \ti e_L)_{T}$ and
$P(\ti e_R \ti e_R)_{T}$ for transversely polarized beams,
but only for longitudinally polarized beams. 
Both are consequences of the Dirac algebra, since
transverse beam polarization is described 
with an additional $\gamma$ matrix;
see Eqs.~(\ref{eq_bouch1-he}) and (\ref{eq_bouch2-he}).

The differential cross section for the process 
$e^+e^-\to\ti\chi^0_i\ti\chi^0_j$ 
is given by
\be{eq:crossectionprod}
{\rm d}\sigma=\frac{1}{2 (2 \pi)^2}
\frac{q}{s^{3/2}}~P(\ti\chi^0_i\ti\chi^0_j)~ {\rm d}\cos\theta~ {\rm d}\phi~,
\ee
where $P(\ti\chi^0_i\ti\chi^0_j)$ contains 
the terms for arbitrary beam polarization and
$q$ is the momentum of the neutralinos in the 
center-of-mass system (cms) 
(see Appendix \ref{appendixA}). 

\subsubsection{CP-behaviour of the kinematical quantities}

In the following, the CP properties of the kinematical
quantity $r_2$, Eq.~\rf{eq:r2}, and 
of the propagator difference $[\Delta(\ti{e},t)-\Delta(\ti{e},u)]$ 
are discussed.
These quantities contribute to the interference
terms of the matrix element squared and are proportional
to the imaginary parts of products of couplings, 
see Eqs.~\rf{eq:PZeL}--\rf{eq:PeLeR}. In the cms,
$r_2$ is given by:
\be{eq:r2defcms}
r_2 = 2 E_b \, [ \vec{t}_+ (\vec{p}_1 \times \vec{p}_4) (t_- p_3)
+ \vec{t}_- (\vec{p}_1 \times \vec{p}_3) (t_+ p_4) ]~,             
\ee
where $E_b$ is the beam energy. Note that the second line of Eq.~\rf{eq:r2}
vanishes in the cms.
Applying a CP transformation to $r_2$, Eq.~\rf{eq:r2defcms}, with
the following transformations 
$(\vec{p}_1,\vec{p}_2,\vec{p}_3,\vec{p}_4,\vec{t}_-)
\stackrel{CP}{\large\longleftrightarrow}
(\vec{p}_1,\vec{p}_2,-\vec{p}_3,-\vec{p}_4,\vec{t}_+)$, we 
find that $r_2$ is CP-even.
Since under CP $\Delta(\ti e,t)\stackrel{CP}{\longleftrightarrow} \Delta(\ti
e,u)$ the propagator differences in Eqs.~\rf{eq:PZeL}--\rf{eq:PeLeR}
are CP-odd, their products with the CP-even quantity $r_2$
are CP-odd.
We emphasize that this
is due to the Majorana nature of the neutralinos, 
which leads to the simultaneous
presence of the $t$- {\it and} $u$-channel contributions,
that the terms in Eqs.~\rf{eq:PZeL}--\rf{eq:PeLeR},
which involve the imaginary part of the couplings, are non-vanishing
in general. 

\subsection{
Contributions to the production spin density matrix 
dependent on neutralino polarization}
We now consider the terms $\Sigma^{b}_P(\ti{\chi}^0_j)$ of the production 
spin density matrix, which 
depend on the polarization 4-vector $s^b$ of the neutralino $\ti\chi^0_j$.
In the following we only list the terms 
$\Sigma^b_P(\ti\chi^0_j)_T$, that 
involve the transverse beam polarization 
(for the contributions independent of the beam polarization and
the terms that depend on the longitudinal beam polarization, see 
\cite{Moortgat-Pick:1999di}):
\begin{equation}
\Sigma^b_P(\ti\chi^0_j)_T=
\Sigma^b_P(Z \tilde{e}_L)_{T}+\Sigma^b_P(Z \tilde{e}_R)_{T}
+\Sigma^b_P(\tilde{e}_L \tilde{e}_R)_{T}~,
\label{eq:defsigbsum}
\end{equation}
\baq{eq:SPZeL}
\Sigma^b_P(Z \tilde{e}_L)_{T}&=& {\cal P}_T^- {\cal P}_T^+~\frac{1}{2}
\frac{g^2}{\cos^2\Theta_W}R_e\nonumber\\
&&{}\times\{{\rm Re}(
\Delta(Z)f^{L*}_{\ell i}f^{L}_{\ell j}O^L_{ij}
[\Delta(\tilde{e}_L,u)^*-\Delta(\tilde{e}_L,t)^*])~r_1^b\nonumber\\
&&{}
-{\rm Im}(
\Delta(Z)f^{L*}_{\ell i}f^{L}_{\ell j}O^L_{ij}
[\Delta(\tilde{e}_L,u)^*+\Delta(\tilde{e}_L,t)^*])~r_2^b\}~,
\eaq
\baq{eq:SPZeR}
\Sigma^b_P(Z \tilde{e}_R)_{T}&=& {\cal P}_T^- {\cal P}_T^+~\frac{1}{2}
\frac{g^2}{\cos^2\Theta_W}L_e\nonumber\\
&&{}\times\{{\rm Re}(
\Delta(Z)f^{R*}_{\ell i}f^{R}_{\ell j}O^R_{ij}
[\Delta(\tilde{e}_R,u)^*-\Delta(\tilde{e}_R,t)^*])~r_1^b\nonumber\\
&&{}
+{\rm Im}(
\Delta(Z)f^{R*}_{\ell i}f^{R}_{\ell j}O^R_{ij}
[\Delta(\tilde{e}_R,u)^*+\Delta(\tilde{e}_R,t)^*])~r_2^b\}~,
\eaq
\baq{eq:SPeLeR}
\Sigma^b_P(\tilde{e}_L \tilde{e}_R)_{T}&=& {\cal P}_T^- {\cal P}_T^+~\frac{1}{4}
g^4 \nonumber\\
&&{}\!\!\!\!\!\!\!\!\!\!\!\!\!
\times \{
{\rm Re}([
\Delta(\tilde{e}_L,t)\Delta(\tilde{e}_R,u)^*-
\Delta(\tilde{e}_L,u)\Delta(\tilde{e}_R,t)^*]
f^{L*}_{\ell i}f^L_{\ell j}f^{R*}_{\ell i}f^R_{\ell j})~r^b_1
\nonumber\\
&&{}\!\!\!\!\!\!\!\!\!\!\!\!\!+
{\rm Im}([
\Delta(\tilde{e}_L,t)\Delta(\tilde{e}_R,u)^*+
\Delta(\tilde{e}_L,u)\Delta(\tilde{e}_R,t)^*]
f^{L*}_{\ell i}f^L_{\ell j}f^{R*}_{\ell i}f^R_{\ell j})~r^b_2\}
\eaq 
with the following notation
\baq{eq:Sr1}
r^b_1 &=&m_{\chi_j}\{[(t_- s^b)(t_+ p_3)+(t_- p_3)(t_+ s^b)]
(p_1 p_2)\nonumber\\
&& +[(p_1 s^b)(p_2 p_3)+(p_1 p_3)(p_2 s^b)-(p_1 p_2)(p_3 s^b)]
(t_- t_+ )\}~,
\eaq
\baq{eq:Sr2}
r^b_2&=& \varepsilon^{\mu\nu\rho\sigma}~m_{\chi_j} [
t_{+,\mu} p_{1\nu} p_{2\rho} s^b_\sigma (t_- p_3) 
+t_{-,\mu} p_{1\nu} p_{2\rho} p_{3\sigma}(t_+ s^b)\nonumber\\
&& 
+ t_{-,\mu} t_{+,\nu} p_{2\rho} s^b_\sigma (p_1 p_3)
+ t_{-,\mu} t_{+,\nu} p_{1\rho} p_{3\sigma} (p_2 s^b)
]~,  
\eaq
where the $e^{\pm}$ polarization vector $t_\pm$ 
is given by Eq.~(\ref{def-trans-basis}).
The polarization basis 4-vectors $s^b$ of the neutralino $\tilde{\chi}^0_j$
fulfil the orthogonality relations $s^b\cdot s^c=-\delta^{bc}$ 
and $s^b\cdot p_4=0$.
The parametrization of the neutralino spin vectors is 
given in Appendix~\ref{appendixA}. The terms 
$\Sigma^a_P(\ti{\chi}^0_i)_T$, which depend on the polarization
4-vector $s^a$ of $\ti{\chi}^0_i$, are obtained by the substitutions  
$s^b\to -s^a, m_{\chi_j} \to m_{\chi_i}, p_3\to p_4$
in Eqs.~\rf{eq:Sr1} and \rf{eq:Sr2}.
Note that, like $P(\ti\chi^0_i\ti\chi^0_j)_T$, 
the expressions $\Sigma^b_P(\ti\chi^0_j)_T$ contain
no contributions 
$\Sigma^b_P(\ti e_L \ti e_L)_T$ and $\Sigma^b_P(\ti e_R \ti e_R)_T$,
but an interference term, $\Sigma^b_P(\ti e_L \ti e_R)_T$.
Furthermore, owing to the Majorana character of the neutralinos,
there is no contribution $\Sigma^b_P(Z Z)_T$.
This is contrary to the cases of unpolarized and
longitudinally polarized beams \cite{Moortgat-Pick:1999di}.

\section{CP asymmetries with transverse beam \\
polarization \label{cpasymmetries}}
\subsection{CP asymmetries in neutralino production}

In this section we construct CP asymmetries for the production process
$e^+e^-\to\ti\chi^0_i\ti\chi^0_j$ 
with transverse $e^+$ and $e^-$ beam polarizations. 
The corresponding cross section is given in Eq.~\rf{eq:crossectionprod}.
Choosing the $e^-$ beam direction along the $z$-axis
in the reference
system (see Appendix \ref{appendixA} and Fig.~\ref{fig_skizze}), the
kinematical quantities in Eqs.~\rf{eq:r1} and \rf{eq:r2} can be rewritten
as
\baq{eq:r1lab}
r_1 &=& -2 E^2_b~q^2\sin^2\theta \cos(\eta-2 \phi)~,
\eaq
\baq{eq:r2lab}
r_2&=& 2 E^2_b~q^2\sin^2\theta \sin(\eta-2 \phi)~,
\eaq
where $\eta=\phi_{-}+\phi_{+}$.
The CP-sensitive terms ($\propto r_2 \propto \sin(\eta-2 \phi)$) can 
be extracted from the amplitude squared 
by an appropriate integration over the azimuthal angle $\phi$.
We define the resulting asymmetry as
\baq{eq:AT1}
A_{CP}(\theta)&=&
\frac{\mbox{N}[\,\sin(\eta-2\phi)>0;\theta\,]
-\mbox{N}[\,\sin(\eta-2\phi)<0;\theta\,]}
{\mbox{N}[\,\sin(\eta-2\phi)>0;\theta\,]
+\mbox{N}[\,\sin(\eta-2\phi)<0;\theta\,]}\nonumber\\[3mm]
&=&\frac{1}{\sigma}
\left[ 
-\int^{\frac{\pi}{2}+\frac{\eta}{2}}_{\frac{\eta}{2}}+
\int^{\pi+\frac{\eta}{2}}_{\frac{\pi}{2}+\frac{\eta}{2}}-
\int^{\frac{3\pi}{2}+\frac{\eta}{2}}_{\pi+\frac{\eta}{2}}+
\int^{2\pi+\frac{\eta}{2}}_{\frac{3\pi}{2}+\frac{\eta}{2}}
\right]
\frac{{\rm d}^2\sigma}{{\rm d}\phi \, {\rm d}\theta} {\rm d}\phi~,
\eaq
which depends on the polar angle $\theta$. The first line
in Eq.~\rf{eq:AT1} exhibits how the asymmetry
is obtained in the experiment, where
$\mbox{N}[\,\sin(\eta-2\phi)> 0 \,(< 0)\,]$
denotes the number of events 
with $\sin(\eta-2\phi)> 0 \,(< 0)$.
The second line in Eq.~\rf{eq:AT1} shows how
the asymmetry is calculated.
We can infer from Eqs.~\rf{eq:PZeL}--\rf{eq:PeLeR} that
$A_{CP}(\theta)$, Eq.~\rf{eq:AT1}, would be zero
if integrated over the whole range of $\theta$: because
of the propagators, the contribution of the $t$-channel
cancels that of the $u$-channel.
Therefore, we divide the integration 
over $\theta$ into two ranges in order to obtain
the CP asymmetry  
\baq{eq:Asy}
A_{CP}&=&
\Big\{\mbox{N}[\,\sin(\eta-2\phi)>0;\cos\theta>0\,]
-\mbox{N}[\,\sin(\eta-2\phi)>0;\cos\theta<0\,]\nonumber\\
&&+\mbox{N}[\,\sin(\eta-2\phi)<0;\cos\theta<0\,]
-\mbox{N}[\,\sin(\eta-2\phi)<0;\cos\theta>0\,]\Big\}/\mathrm{N_{tot}}
\nonumber\\[3mm]
&=&\left[\int^{\pi/2}_{0} -\int^{\pi}_{\pi/2}\right] 
A_{CP}(\theta)\,{\rm d}\theta~,
\eaq
where $\mathrm{N_{tot}}$ denotes the total number of events. 
Note that for a measurement of the CP asymmetry in Eq.~\rf{eq:Asy} the 
production plane has to be reconstructed. In 
Appendix \ref{appendixC} we propose how this can be done. 

Finally we remark that an azimuthal asymmetry, analogous
to that studied for chargino production \cite{Bartl:2004xy}, can be defined
also for neutralino production. 
It is given by
\baq{eq:Aphi}
A_{\phi}&=&
\frac{\mbox{N}[\,\cos(\eta-2\phi)>0\,]
-\mbox{N}[\,\cos(\eta-2\phi)<0\,]}
{\mbox{N}[\,\cos(\eta-2\phi)>0\,]
+\mbox{N}[\,\cos(\eta-2\phi)<0\,]}\nonumber\\[3mm]
&=&\frac{1}{\sigma}
\left[
-\int^{\frac{3\pi}{4}+\frac{\eta}{2}}_{\frac{\pi}{4}+\frac{\eta}{2}}+
\int^{\frac{5\pi}{4}+\frac{\eta}{2}}_{\frac{3\pi}{4}+\frac{\eta}{2}}-
\int^{\frac{7\pi}{4}+\frac{\eta}{2}}_{\frac{5\pi}{4}+\frac{\eta}{2}}+
\int^{\frac{9\pi}{4}+\frac{\eta}{2}}_{\frac{7\pi}{4}+\frac{\eta}{2}}
\right]
\frac{{\rm d}\sigma}{{\rm d}\phi} \, {\rm d}\phi~.
\eaq
In this case the integration over the polar 
angle $\theta$ is performed over the whole range. 
This choice of the ranges of the integrations 
has the effect of extracting the terms $\propto r_1\propto 
\cos(\eta-2\phi)$, Eqs.~\rf{eq:PZZ}--\rf{eq:PeLeR}, 
from the squared amplitude. 
Note however, that this observable is CP-even.

\subsection{CP asymmetries in neutralino production and decay} 

The reconstruction of the neutralino momenta is not 
necessary if we include the subsequent decays 
$\ti{\chi}^0_j \to \ti\ell^\pm~\ell_1^\mp$ 
(where $\ti\ell=\ti\ell_L,\ti\ell_R$) and
$\ti{\ell}^\pm\to\ell^\pm_2\ti{\chi}^0_1$, yielding
to the final state $\ti{\chi}^0_j \to\ell^\mp_1\ell^\pm_2\ti\chi^0_1$.
The label of the leptons indicates whether they
stem from the first or the second decay.
The cross sections for the combined processes are given 
in Appendix \ref{appendixB}, Eqs.~\rf{eq:crossection} and 
\rf{eq:crossection2}, respectively.
The CP-sensitive terms of the squared amplitudes depend on
$\sin (\eta-2\phi_{\ell_{1}})$ or $\sin (\eta-2\phi_{\ell_{2}})$, where
$\phi_{\ell_1}$ and $\phi_{\ell_2}$ are the azimuthal
angles of the final leptons $\ell_1^\mp$ and $\ell_2^\pm$.
As a first step, we integrate the differential cross section in 
Eq.~\rf{eq:crossection}
over all angles except $\phi_{\ell_{1}}$ 
(the angles are integrated over their whole range). 
Then the CP asymmetry obtained by the azimuthal distribution of $\ell^-_1$ 
is given by 
\baq{eq:AT2}
A_{1}^-&=&
\frac{\mbox{N}[\,\sin(\eta-2\phi_{\ell_1})>0\,]
-\mbox{N}[\,\sin(\eta-2\phi_{\ell_1})<0\,]}
{\mbox{N}[\,\sin(\eta-2\phi_{\ell_1})>0\,]
+\mbox{N}[\,\sin(\eta-2\phi_{\ell_1})<0\,]}\nonumber\\[3mm]
&=&\frac{1}{\sigma_1}
\left[
-\int^{\frac{\pi}{2}+\frac{\eta}{2}}_{\frac{\eta}{2}}+
\int^{\pi+\frac{\eta}{2}}_{\frac{\pi}{2}+\frac{\eta}{2}}-
\int^{\frac{3\pi}{2}+\frac{\eta}{2}}_{\pi+\frac{\eta}{2}}+
\int^{2\pi+\frac{\eta}{2}}_{\frac{3\pi}{2}+\frac{\eta}{2}}
\right]
\frac{{\rm d}\sigma_1}{{\rm d}\phi_{\ell_1}} {\rm d}\phi_{\ell_1}~,
\eaq
where 
$\sigma_1=\sigma(e^+e^-\to\ti\chi^0_1\ti\chi^0_j)\times
B(\ti\chi^0_j\to\ti\ell^+ \ell_1^-)$ 
and the upper index of $A_{1}^-$ corresponds
to the electric charge of the observed lepton $\ell^-_1$. 

As a next step, we integrate the differential cross section in 
Eq.~\rf{eq:crossection2}
over all angles except $\phi_{\ell_{2}}$, in order to define
the CP asymmetry of the azimuthal distribution of $\ell^+_2$:
\baq{eq:AT21}
A_2^+&=&
\frac{\mbox{N}[\,\sin(\eta-2\phi_{\ell_2})>0\,]
-\mbox{N}[\,\sin(\eta-2\phi_{\ell_2})<0\,]}
{\mbox{N}[\,\sin(\eta-2\phi_{\ell_2})>0\,]
+\mbox{N}[\,\sin(\eta-2\phi_{\ell_2})<0\,]}\nonumber\\[3mm]
&=&\frac{1}{\sigma_2}
\left[
-\int^{\frac{\pi}{2}+\frac{\eta}{2}}_{\frac{\eta}{2}}+
\int^{\pi+\frac{\eta}{2}}_{\frac{\pi}{2}+\frac{\eta}{2}}-
\int^{\frac{3\pi}{2}+\frac{\eta}{2}}_{\pi+\frac{\eta}{2}}+
\int^{2\pi+\frac{\eta}{2}}_{\frac{3\pi}{2}+\frac{\eta}{2}}
\right]
\frac{{\rm d}\sigma_2}{{\rm d}\phi_{\ell_2}} {\rm d}\phi_{\ell_2}~,
\eaq
where
$\sigma_2=\sigma(e^+e^-\to\ti\chi^0_1\ti\chi^0_j)\times
B(\ti\chi^0_j\to\ti\ell^+ \ell^-_1)\times B(\ti\ell^+\to\ti\chi^0_1 \ell^+_2)$.
Note that, since $\Sigma^b_D(\ti{\chi}^0_j)$ 
for the two C-conjugate decay modes of
$\ti\chi_j^0\to\ti\ell^\pm\ell^\mp$ differs only by 
a sign (see Eqs.~\rf{eq:SDslepL} and \rf{eq:SDslepR})
the asymmetries with upper indices  $+$ and $-$ are related by
$A_i^+=-A_i^-,~i=1,2$.
In order to measure both 
asymmetries, Eqs.~\rf{eq:AT2} and \rf{eq:AT21}, 
it is necessary to distinguish the lepton
$\ell^{\mp}_1$, originating from the decay $\ti\chi^0_j\to \ti \ell^{\pm}
\ell_1^{\mp}$, and the lepton $\ell^{\pm}_2$ from the subsequent decay
$\ti \ell^\pm\to \ti\chi^0_1 \ell^{\pm}_2$. This can be accomplished 
by their different energy distributions, 
when the masses of the particles involved are known,
provided that their measured energies do not 
lie in the overlapping region of their energy distributions
\cite{Bartl:2003tr}. 

However, we can also define a CP asymmetry where 
it is not necessary to distinguish whether the leptons 
stem from the first or the second 
step of the decay chain $\ti{\chi}^0_j \to \ti\ell^\pm~\ell_1^\mp 
\to\ell^\mp_1\ell^\pm_2\ti\chi^0_1$. 
This asymmetry is defined by
\baq{eq:ATp}
A^- &=&
\frac{\mbox{N}[\,\sin(\eta-2\phi_{\ell^-})>0\,]
-\mbox{N}[\,\sin(\eta-2\phi_{\ell^-})<0\,]}
{\mbox{N}[\,\sin(\eta-2\phi_{\ell^-})>0\,]
+\mbox{N}[\,\sin(\eta-2\phi_{\ell^-})<0\,]}
\nonumber\\[3mm]
{}&=&
\frac{
(\int^+-\int^-)(
\frac{{\rm d}\sigma_1}{{\rm d}\phi_{\ell_1}} {\rm d}\phi_{\ell_1}+
\frac{{\rm d}\sigma_2}{{\rm d}\phi_{\ell_2}}{\rm d}\phi_{\ell_2})
}
{\int^{2\pi}_0(\frac{{\rm d}\sigma_1}{{\rm d}\phi_{\ell_1}}{\rm
d}\phi_{\ell_1}+
\frac{{\rm d}\sigma_2}{{\rm d}\phi_{\ell_2}} {\rm d}\phi_{\ell_2})}~,
\eaq
where $\ell^-$ is either $\ell^-_1$ or $\ell^-_2$
and $\mbox{N}[\,\sin(\eta-2\phi_{\ell^-})> 0\,(< 0)\,]$
denotes the number of events 
where $\sin(\eta-2\phi_{\ell^-})> 0\,(< 0)$.
Hence, only the charge of the lepton and its azimuthal
angle $\phi_{\ell^-}$ has to be determined.
In Eq.~\rf{eq:ATp} $\int^{\pm}$ corresponds to 
an integration over the azimuthal
angles $\phi_{\ell_1}$ or $\phi_{\ell_2}$, where
$\sin(\eta-2\phi_{\ell_{1,2}})$ is positive or negative, respectively.
An analogous asymmetry can be defined for $\ell^+$ as well.
The asymmetry $A^-$, Eq.~\rf{eq:ATp}, can be related to the asymmetries
$A^-_1$ and $A^-_2$, Eqs.~\rf{eq:AT2} and \rf{eq:AT21}, by
\be{eq:ATrel}
A^-=
\frac{1}{[1+B(\ti\ell^-\to\ell^-\ti\chi_1^0)]}
\left[A^-_1+A^-_2~B(\ti\ell^-\to\ell^-\ti\chi_1^0)\right]~.
\ee
%

\section{Numerical studies \label{numerics}}

\subsection{CP-even observables in neutralino production}
Before we concentrate on the numerical study of CP-odd observables, 
we would like to give an example, which shows
that a measurement of only CP-even observables may not
be sufficient to unambiguously determine the SUSY parameters
of the neutralino sector. However, a measurement of a 
CP-odd asymmetry may help to single out the correct solution.
This may be particularly important if only the two lower states
of the neutralino spectrum are kinematically accessible. 
\begin{table}[H]
\begin{center}
\begin{tabular}{|c||c|c|c|c|c|c|c|c|} \hline
 Scenario 
& \multicolumn{1}{c|}{$|M_1|$} 
& \multicolumn{1}{c|}{$\phi_{M_1}$} 
& \multicolumn{1}{c|}{$M_2$} 
& \multicolumn{1}{c|}{$|\,\mu\,|$}
& \multicolumn{1}{c|}{$\phi_{\mu}$}  
& \multicolumn{1}{c|}{$\tan{\beta}$}  
& \multicolumn{1}{c|}{$m_{\tilde{e}_L}$}
& \multicolumn{1}{c|}{$m_{\tilde{e}_R}$}\\\hline\hline
 Complex 
& \multicolumn{1}{c}{183} 
& \multicolumn{1}{c}{$0.05\pi$} 
& \multicolumn{1}{c}{311} 
& \multicolumn{1}{c}{343}  
& \multicolumn{1}{c}{$1.9\pi$}
& \multicolumn{1}{c}{2.1} 
& \multicolumn{1}{c}{297}
& \multicolumn{1}{c|}{181}\\\hline 
 Real 
& \multicolumn{1}{c}{180 } 
& \multicolumn{1}{c}{0} 
& \multicolumn{1}{c}{310} 
& \multicolumn{1}{c}{335} 
& \multicolumn{1}{c}{0}
& \multicolumn{1}{c}{3}  
& \multicolumn{1}{c}{300}
& \multicolumn{1}{c|}{180}\\\hline
\end{tabular}\\[0.5ex]
\caption{\label{taboddeven}
Input parameters $|M_1|$, $\phi_{M_1}$, $M_2$, $|\mu|$, 
$m_{\tilde{e}_L}$ and $m_{\tilde{e}_R}$
for the complex and the real scenario.
All mass parameters are given in GeV.}
\end{center}
\end{table}
To this end we consider the complex scenario with the 
parameters given in Table~\ref{taboddeven},
leading to $m_{\ti{\chi}^0_1}=170.9$~GeV and $m_{\ti{\chi}^0_2}=259.5$~GeV.
At $\sqrt{s}=500$~GeV only the cross sections of 
$e^+e^- \to \ti{\chi}^0_1 \ti{\chi}^0_2$ would be
measurable, giving 
$\sigma(e^+e^- \to \ti{\chi}^0_1 \ti{\chi}^0_2)=(16.4,18.3,30.3)$~fb
for the $e^+$ and $e^-$ longitudinal beam polarizations 
$({\cal P}^-_L,{\cal P}^+_L)=(0,0)$, $(-80\%,+60\%)$, $(+80\%,-60\%)$,
respectively. We assume that the masses $m_{\ti{\chi}^0_{1,2}}$ and
$m_{\ti{e}_{L,R}}$ are measured with $1\%$ accuracy. For the cross
sections we take an error corresponding to a 
1-$\sigma$ deviation for a luminosity 
$\mathcal{L}_{\rm int}=100~\mbox{fb}^{-1}$.
Then within this accuracy, we would obtain compatible neutralino
masses and cross sections with the real SUSY parameter set, 
which is also given in Table~\ref{taboddeven}, 
namely $m_{\ti{\chi}^0_1}=169.3$~GeV 
and $m_{\ti{\chi}^0_2}=258.3$~GeV, and cross sections
$\sigma(e^+e^- \to \ti{\chi}^0_1 \ti{\chi}^0_2)=(16.3,18.2,30.0)$~fb.  
The CP-odd
asymmetry ${\cal A}_{CP}$, Eq.~\rf{eq:Asy}, 
however, would result in about $2.8\%$ with 
$({\cal P}^-_T,{\cal P}^+_T)=(80\%,60\%)$ for the complex
scenario with $\phi_{M_1}=0.05\pi$ and $\phi_{\mu}=1.9\pi$. 
Although the asymmetry is small it should be experimentally
measurable including the statistical uncertainty. Therefore
the complex scenario would be clearly distinguishable
from the real scenario, which results in an asymmetry 
identical to zero.
This simple example
illustrates that it is necessary to measure CP-odd observables
for truly identifying CP-violating effects.

In the following we analyse numerically the CP-odd asymmetries,
Eq.~\rf{eq:Asy} and Eqs.~\rf{eq:AT2}--\rf{eq:ATp}, 
at the ILC
with $\sqrt{s}=500$~GeV and transversely polarized $e^\pm$ beams.
We especially focus on the influence of the phase $\phi_{M_1}$
of the gaugino mass parameter $M_1=|M_1|e^{i\phi_{M_1}}$. 
Throughout we assume the GUT-inspired releation 
$|M_1|=5/3\tan^2\Theta_W \, M_2$. Furthermore
we show that CP-odd observables are necessary to determine
unambiguously the underlying SUSY parameters. In order
to study the full phase dependences of the CP-odd observables,
we do not take into account the restrictions from the EDMs
and vary $\phi_\mu$ and $\phi_{M_1}$ in the whole range.
 
\subsection{CP-odd asymmetries in neutralino production}
\begin{figure}[t]
\setlength{\unitlength}{1mm}
\begin{center}
\begin{picture}(150,35)
\put(-53,-150){\mbox{\epsfig{figure=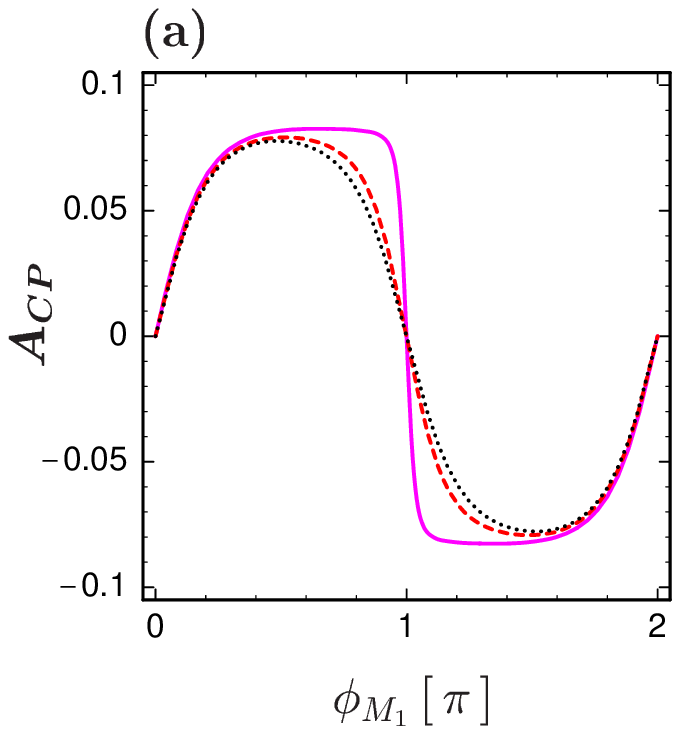,height=22.cm,width=19.4cm}}}
\put(27,-150){\mbox{\epsfig{figure=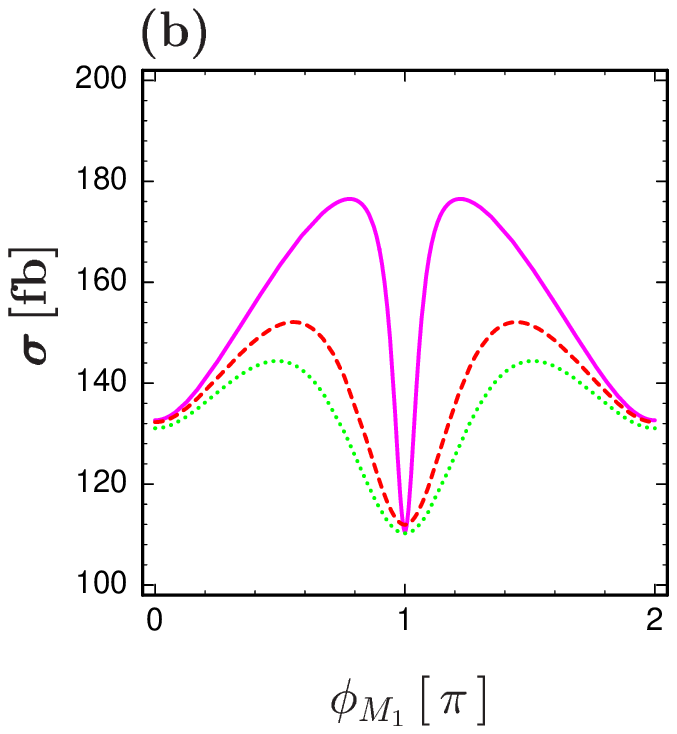,height=22.cm,width=19.4cm}}}
\put(27,44){\small ${\bf e^+e^-\to \tilde{\chi}^0_1\tilde{\chi}^0_2}$}
\put(107,44){\small ${\bf e^+e^-\to \tilde{\chi}^0_1\tilde{\chi}^0_2}$}
\end{picture}
\end{center}
\caption{(a) CP asymmetry $A_{CP}$, Eq.~\rf{eq:Asy}, and (b)
cross section $\sigma(e^+e^- \to \ti{\chi}^0_1 \ti{\chi}^0_2)$
as a function of $\phi_{M_1}$ for scenario A of Tab.\ref{scentab},
with $\tan\beta=3$ (solid line),
$\tan\beta=10$ (dashed line), $\tan\beta=30$ (dotted line),
for $\sqrt{s}=500$~GeV and 
transverse beam polarizations $({\cal P}_T^-, {\cal P}_T^+)=(100\%,100\%)$. }
\label{fig:plot1}
\end{figure}

First we discuss the CP-odd asymmetry $A_{CP}$, Eq.~\rf{eq:Asy},
for the neutralino production processes 
$e^+e^- \to \ti{\chi}^0_1 \ti{\chi}^0_2$
and $e^+e^- \to \ti{\chi}^0_1 \ti{\chi}^0_3$.
CP violation is due
to the interference terms $P(Z \tilde{e}_L)_{T}$, $P(Z \tilde{e}_R)_{T}$,
and $P(\tilde{e}_L \tilde{e}_R)_{T}$, Eqs.~\rf{eq:PZeL}--\rf{eq:PeLeR}. 
We assume that the momenta of the produced
neutralinos can be reconstructed by analysing
the subsequent two-body decays; see Appendix~\ref{appendixC}.

\begin{table}[t]
\begin{center}
\begin{tabular}{|c||c|c|c|} \hline
 Scenario & \multicolumn{2}{c|}{A} & \multicolumn{1}{c|}{B}\\
\hline\hline
 $|M_1|$ & \multicolumn{2}{c|}{123.3}  
& \multicolumn{1}{c|}{120.8}\\ \hline
$\phi_{M_1}$ & \multicolumn{2}{c|}{$0.5\pi$} 
& \multicolumn{1}{c|}{$0.5\pi$}\\ \hline
 $M_2$ & \multicolumn{2}{c|}{245}  
& \multicolumn{1}{c|}{240}\\ \hline
 $|\,\mu\,|$ & \multicolumn{2}{c|}{160} 
& \multicolumn{1}{c|}{300}\\ \hline
$\phi_{\mu}$ & \multicolumn{2}{c|}{$0$} 
& \multicolumn{1}{c|}{$0$}\\ \hline
 $m_{\tilde{e}_L}$ & \multicolumn{2}{c|}{400} 
& \multicolumn{1}{c|}{400} 
\\ \hline
 $m_{\tilde{e}_R}$ & \multicolumn{2}{c|}{150}  
& \multicolumn{1}{c|}{150}\\
\hline\hline
 $\tan{\beta}$ & $3$  & $30$  
& $3$\\ \hline
 $m_{\tilde{\chi}^0_1}$ & 99.4 & 105.5 & 117.0\\ \hline
 $m_{\tilde{\chi}^0_2}$ & 143.0 & 144.1 & 197.6\\ \hline
 $m_{\tilde{\chi}^0_3}$ & 169.7 & 178.6 & 303.9\\ \hline
 $m_{\tilde{\chi}^0_4}$ & 289.7 & 281.5 & 351.7\\ \hline
\end{tabular}\\[0.5ex]
\caption{\label{scentab}
Input parameters $|M_1|$, $M_2$, $|\mu|$, 
$m_{\tilde{e}_L}$ and $m_{\tilde{e}_R}$
and the resulting masses $m_{\tilde{\chi}^0_i}$, $i=1,\ldots,4$ for
$\tan{\beta}=3,30$ and specific values of the phases $\phi_{M_1}$ and 
$\phi_\mu$. All mass parameters and masses are given in GeV.}
\end{center}
\end{table}

\subsubsection{CP-odd asymmetries in 
${\bf e^+e^-\to\tilde{\chi}^0_1\tilde{\chi}^0_2}$ production}

In Fig.~\ref{fig:plot1}a we show $A_{CP}$, Eq.~\rf{eq:Asy}, for 
$e^+e^- \to \ti{\chi}^0_1 \ti{\chi}^0_2$
as a function of $\phi_{M_1}$ for scenario A, given
in Table~\ref{scentab}, for $\tan\beta=3,10,30$, with
$\sqrt{s}=500$~GeV and transverse beam polarization 
$({\cal P}_T^-, {\cal P}_T^+)=(100\%,100\%)$.  
For this scenario we obtain for $\tan\beta=3$ $(30)$ 
an asymmetry $A_{CP}$ of
about $8.2$ $(7.8)$\%, for $\phi_{M_1}=0.5\pi$.
The peculiar shape of the curve is a result of
the combined contributions of the $Z$--$\tilde{e}_R$
interference term, which has its maximum at $\phi_{M_1}\approx 0.4\pi$,
and of the $\tilde{e}_L$--$\tilde{e}_R$ interference term, 
with its maximum at $\phi_{M_1}\approx 0.8\pi$. The contribution
of the $Z$--$\tilde{e}_L$ interference term  
is suppressed because of the large mass of the left-handed selectron.
The cross sections for the process 
$e^+e^- \to \ti{\chi}^0_1 \ti{\chi}^0_2$
are plotted in Fig.~\ref{fig:plot1}b and are
about $163$ $(144)$~fb for $\phi_{M_1}=0.5\pi$.
Note that the cross sections are independent 
of the transverse beam polarization,
because these contributions depend on $\cos 2\phi$ $(\sin 2\phi)$, 
see Eq.~\rf{eq:r1lab},
and disappear if integrated over the whole range of $\phi$.
In Figs.~\ref{fig:plot1}a and b we can clearly 
see the antisymmetric dependence of 
the CP asymmetry and the symmetric behaviour of the cross section 
on the phase $\phi_{M_1}$.
It is therefore obvious that  
both kinds of observables are needed for an unambiguous determination of
the phase. Note that $A_{CP}$ can be sizeable even for values
of $\phi_{M_1}$ close to $0$ and $\pi$, which would be favoured by the
EDM constraints.
\begin{figure}[t]
\setlength{\unitlength}{1mm}
\begin{center}
\begin{picture}(150,120)
\put(-53,-110){\mbox{\epsfig{figure=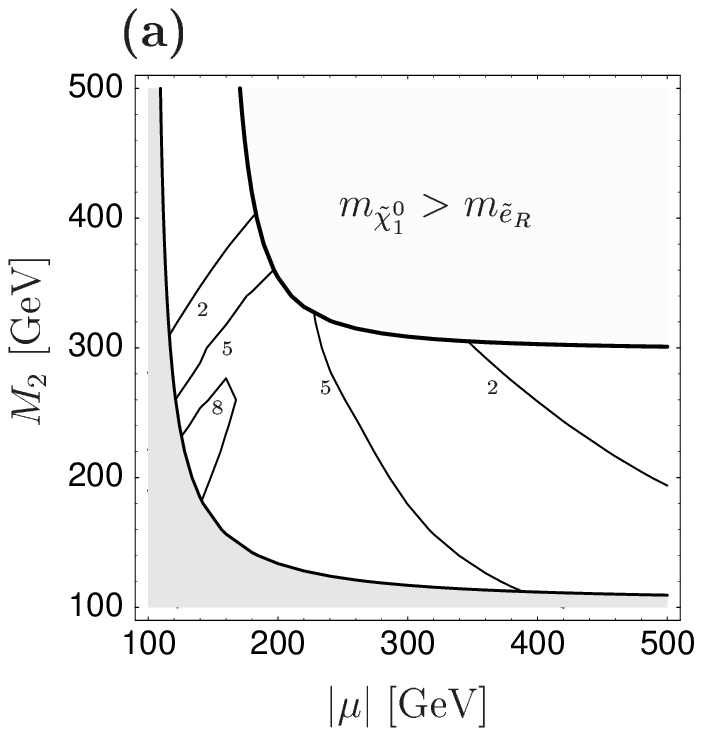,height=27cm,width=19.4cm}}}
\put(27,-110){\mbox{\epsfig{figure=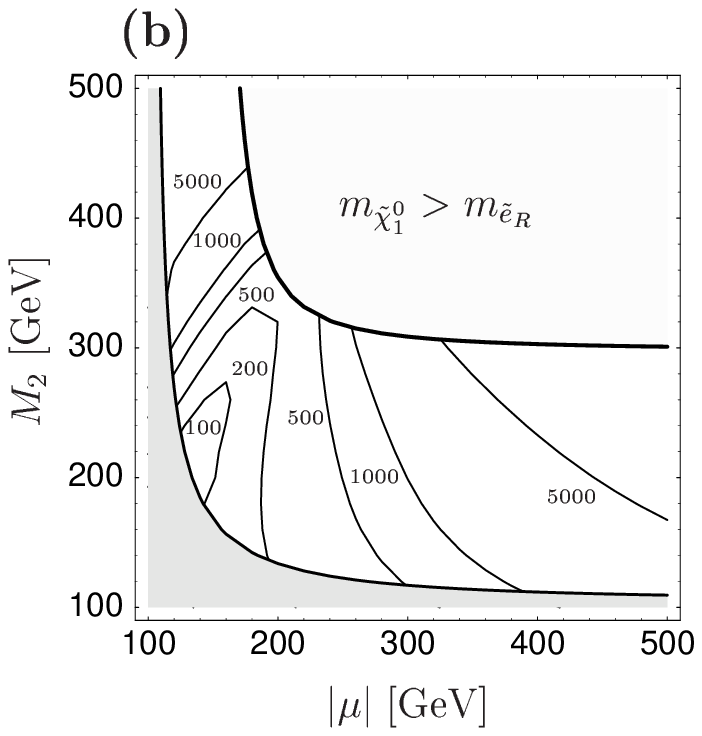,height=27cm,width=19.4cm}}}
\put(17,129){\small ${\bf A_{CP}\,[\%]\,\,\mbox{for} \,\,
e^+e^- \to \ti{\chi}^0_1 \ti{\chi}^0_2}$}
\put(97,129){\small ${\bf \mathcal{L}_{\rm int}\,[\mbox{fb}^{-1}]\,\,
\mbox{for} \,\,
e^+e^- \to \ti{\chi}^0_1 \ti{\chi}^0_2}$}
\end{picture}
\end{center}
\vspace{-7cm}
\caption{(a) contours  
of the CP asymmetry $A_{CP}$, Eq.~\rf{eq:Asy}, in \% 
for the process $e^+e^- \to \ti{\chi}^0_1 \ti{\chi}^0_2$
in the $|\mu|$--$M_2$ plane.
The MSSM parameters are 
$\phi_{M_1}=0.5\pi$, $\phi_{\mu}=0$,
$\tan\beta=3$, $m_{\ti{e}_L}=400$~GeV and $m_{\ti{e}_R}=150$~GeV 
at $\sqrt{s}=500$~GeV with transverse beam polarizations 
$({\cal P}_T^-, {\cal P}_T^+)=(100\%,100\%)$.
(b) shows the contours of the 
luminosity $\mathcal{L}_{\rm int}$, Eq.~\rf{eq:estimate},
needed to measure the CP-odd asymmetry $A_{CP}$
at the 5-$\sigma$ level with 
degrees of transverse polarization $({\cal P}_T^-, {\cal P}_T^+)=(80\%,60\%)$.
The light-grey region is experimentally excluded 
by the exclusion bound $m_{\ti{\chi}^\pm_1}<104$~GeV \cite{LEP}.
}
\label{fig:plot2}
\end{figure}

Now we estimate the observability of the asymmetry.
One assumes that the same degree of transverse beam polarization
is feasible as for the longitudinal polarization 
(${\cal P}_T^-=80$\% and ${\cal P}_T^+=60$\%).
Since the CP asymmetry $A_{CP}$ depends bilinearly on the degrees of
transverse beam polarization ${\cal P}_T^-$~$({\cal P}_T^+)$ of the
$e^-$~$(e^+)$, see Eqs.~\rf{eq:PZeL}--\rf{eq:PeLeR},
we have to multiply the asymmetry $A_{CP}$ for 
$({\cal P}_T^-, {\cal P}_T^+)=(100\%,100\%)$ with a factor $0.48$. 
The luminosity $\mathcal{L}_{\rm int}$ required
for a measurement with specific significance can be estimated
as 
\be{eq:estimate} 
\mathcal{L}_{\rm int}=(\mathcal{N}_{\sigma})^2/[A_{CP}^2 \,\sigma]~,
\ee
where $\mathcal{N}_{\sigma}$ denotes the number of standard deviations
and $\sigma$ the corresponding cross section for neutralino production.
We obtain a luminosity 
$\mathcal{L}_{\rm int} \approx 99$~$(124)~\mbox{fb}^{-1}$ 
needed for a discovery with 5-$\sigma$, 
for $\tan\beta=3$~$(30)$ and $\phi_{M_1}=0.5\pi$.

\begin{figure}[t]
\setlength{\unitlength}{1mm}
\begin{center}
\begin{picture}(150,120)
\put(-53,-110){\mbox{\epsfig{figure=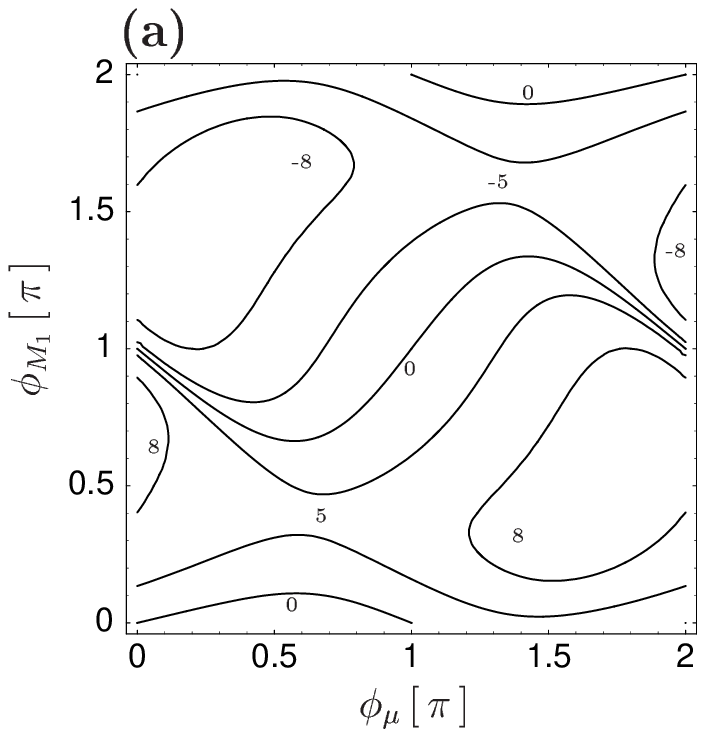,height=27cm,width=19.4cm}}}
\put(27,-110){\mbox{\epsfig{figure=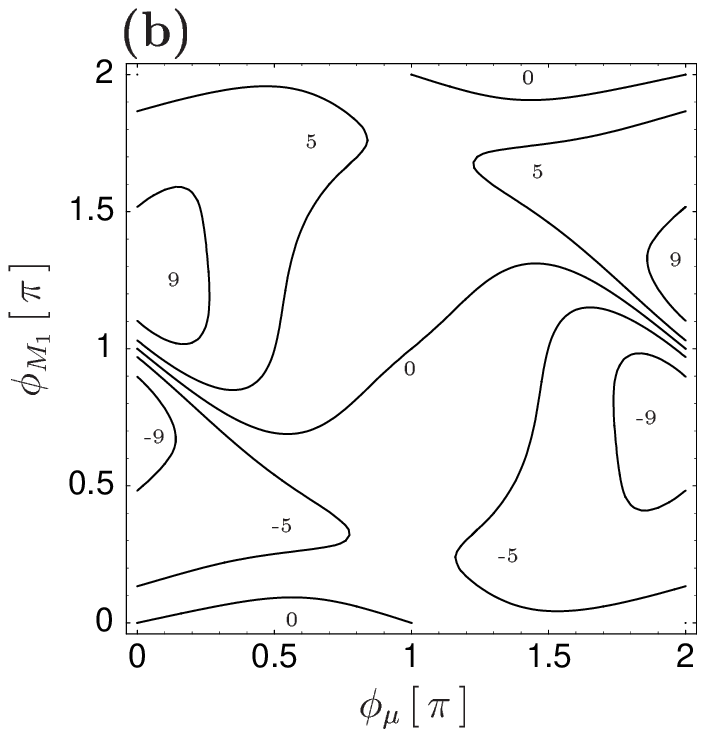,height=27cm,width=19.4cm}}}
\put(17,129){\small ${\bf A_{CP}\,[\%]\,\,\mbox{for} \,\,
e^+e^- \to \ti{\chi}^0_1 \ti{\chi}^0_2}$}
\put(98,129){\small ${\bf A_{CP}\,[\%]\,\,\mbox{for} \,\,
e^+e^- \to \ti{\chi}^0_1 \ti{\chi}^0_3}$}
\end{picture}
\end{center}
\vspace{-7cm}
\caption{Contours  
of the CP asymmetry $A_{CP}$, Eq.~\rf{eq:Asy}, in \% 
for the process (a)
$e^+e^- \to \ti{\chi}^0_1 \ti{\chi}^0_2$
and (b) $e^+e^- \to \ti{\chi}^0_1 \ti{\chi}^0_3$
in the $\phi_\mu$--$\phi_{M_1}$ plane, for scenario A with $\tan\beta=3$, 
see Table~\ref{scentab}, 
for $\sqrt{s}=500$~GeV and
transverse beam polarizations 
$({\cal P}_T^-, {\cal P}_T^+)=(100\%,100\%)$.
}
\label{fig:plot7}
\end{figure} 

Figure~\ref{fig:plot2}a shows the contour lines of the
CP asymmetry $A_{CP}$, Eq.~\rf{eq:Asy}, at $\sqrt{s}=500$~GeV for 
$e^+e^- \to \ti{\chi}^0_1 \ti{\chi}^0_2$
in the $|\mu|$--$M_2$ plane. The MSSM parameters are chosen 
to be $\phi_{M_1}=0.5\pi$, $\phi_{\mu}=0$,
$\tan\beta=3$, $m_{\ti{e}_L}=400$~GeV and $m_{\ti{e}_R}=150$~GeV.
The largest CP-odd asymmetry $A_{CP}$ is attained for 
sizeable gaugino--higgsino mixing.
If the beams are fully transversely polarized,
$({\cal P}_T^-, {\cal P}_T^+)=(100\%,100\%)$, then
$A_{CP}$ could reach up to about $8.8\%$ for 
$M_2 \approx 240$~GeV and $|\mu|\approx 140$~GeV.
With a higher centre-of-mass energy $\sqrt{s}=800$~GeV,
the asymmetry $A_{CP}$ increases to about $12\%$, because
the cross section, which is the denominator of $A_{CP}$, 
decreases stronger than its numerator. 
In this region of the parameter space the $Z$--$\tilde{e}_R$
interference term, Eq.~\rf{eq:PZeR},
is the main contribution to the asymmetry $A_{CP}$.
In a gaugino-like scenario, for instance $M_2=250$~GeV and
$|\mu|=450$~GeV, the $\tilde{e}_L$--$\tilde{e}_R$ interference term
is dominant and the others are suppressed. Generally the 
$\tilde{e}_L$--$\tilde{e}_R$ contribution to the asymmetry is small
for $e^+e^- \to \ti{\chi}^0_1 \ti{\chi}^0_2$ and 
$A_{CP}$ is therefore reduced to about $1.6\%$.
In order to obtain a larger
$\tilde{e}_L$--$\tilde{e}_R$ contribution, a larger mass
splitting of $\ti{e}_L$ and $\ti{e}_R$ is necessary.
If $m_{\ti{e}_L}\approx m_{\ti{e}_R}$ 
the interference term $P(\ti{e}_L\ti{e}_R)_T$ is very small, 
see Eq.~\rf{eq:PeLeR}.
In Fig.~\ref{fig:plot2}b
we plot the corresponding luminosity $\mathcal{L}_{\rm int}$,
Eq.~\rf{eq:estimate}, for transverse beam
polarizations of $({\cal P}_T^-, {\cal P}_T^+)=(80\%,60\%)$. For the maximum
value of $A_{CP}$, 
a luminosity $\mathcal{L}_{\rm int}$ of about $81~\mbox{fb}^{-1}$ would be needed 
for a discovery with 5-$\sigma$.

In Fig.~\ref{fig:plot7}a we show the contour lines of $A_{CP}$, 
Eq.~\rf{eq:Asy}, for $e^+e^- \to \ti{\chi}^0_1 \ti{\chi}^0_2$
in the $\phi_\mu$--$\phi_{M_1}$ plane for scenario A (Table~\ref{scentab})
at $\sqrt{s}=500$~GeV and with transverse beam polarizations 
$({\cal P}_T^-, {\cal P}_T^+)=(100\%,100\%)$. We obtain 
a maximum value of the
CP-odd asymmetry $A_{CP}$ of about $8.9\%$ for 
$\phi_\mu \approx 1.6\pi$ and 
$\phi_{M_1} \approx 0.4\pi$.
In this scenario the $\phi_{M_1}$ and the 
$\phi_\mu$ dependence are of the same order 
of magnitude. The main contribution
to the CP-odd asymmetry originates from the
interference term $P(Z \tilde{e}_R)_{T}$,
Eq.~\rf{eq:PZeR}, i.e.\ the primarily involved coupling
is $f^{R*}_{\ell 1}f^{R}_{\ell 2}O^R_{12}$. 
The corresponding cross
section for $\phi_\mu = 1.6\pi$ and 
$\phi_{M_1} = 0.4\pi$ is about $139$~fb and the 
luminosity $\mathcal{L}_{\rm int}$ for a discovery with 5-$\sigma$  
is about $99~\mbox{fb}^{-1}$ for transverse beam
polarizations $({\cal P}_T^-, {\cal P}_T^+)=(80\%,60\%)$.
Note also in this case the CP asymmetry $A_{CP}$ 
can be sizeable for values of $\phi_{M_1}$ and $\phi_\mu$ 
close to $0$ and $\pi$. 

\subsubsection{CP-odd asymmetries in 
${\bf e^+e^-\to\tilde{\chi}^0_1\tilde{\chi}^0_3}$ production}

Figure~\ref{fig:plot7}b shows the contour lines of the
CP asymmetry $A_{CP}$, Eq.~\rf{eq:Asy}, in the $\phi_\mu$--$\phi_{M_1}$ plane
for $e^+e^- \to \ti{\chi}^0_1 \ti{\chi}^0_3$.
As shown in the case before, a large gaugino--higgsino mixing
is necessary to obtain sizeable CP asymmetries.
We investigate scenario A of Table~\ref{scentab}, at $\sqrt{s}=500$~GeV
and $({\cal P}_T^-, {\cal P}_T^+)=(100\%,100\%)$. 
In this scenario the maximum value of $A_{CP}$
is about $9.8\%$ for $\phi_\mu \approx 0.1\pi$ and
$\phi_{M_1} \approx 1.2\pi$. 
Again the main CP-violating contribution is due to the
$Z$--$\ti{e}_R$ interference term.
In this example 
the largest asymmetries are obtained for small
values of $\phi_\mu$.
For $\phi_\mu = 0.1\pi$ and $\phi_{M_1} = 1.2\pi$ 
the cross
section $\sigma(e^+e^- \to \ti{\chi}^0_1 \ti{\chi}^0_3)$ 
is $76$~fb and the luminosity for a discovery with 5-$\sigma$
is about $150~\mbox{fb}^{-1}$. 

\begin{figure}[t]
\setlength{\unitlength}{1mm}
\begin{center}
\begin{picture}(150,35)
\put(-53,-150){\mbox{\epsfig{figure=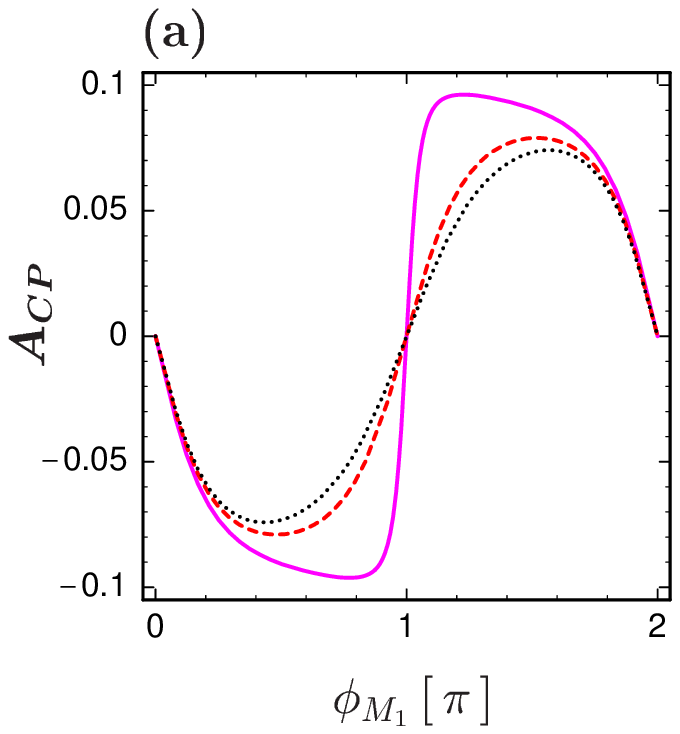,height=22.cm,width=19.4cm}}}
\put(27,-150){\mbox{\epsfig{figure=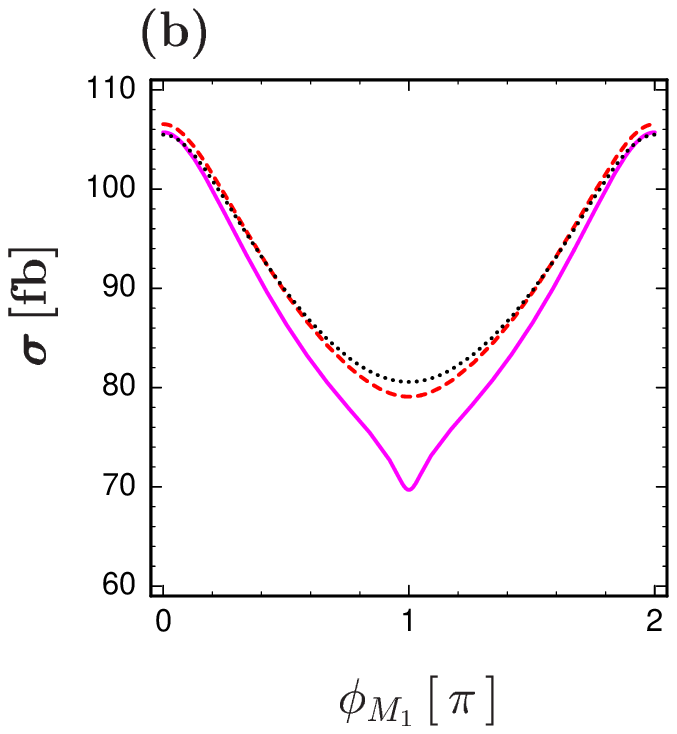,height=22.cm,width=19.4cm}}}
\put(27,44){\small ${\bf e^+e^-\to \tilde{\chi}^0_1\tilde{\chi}^0_3}$}
\put(107,44){\small ${\bf e^+e^-\to \tilde{\chi}^0_1\tilde{\chi}^0_3}$}
\end{picture}
\end{center}
\caption{(a) CP asymmetry $A_{CP}$, Eq.~\rf{eq:Asy}, and (b)
cross section $\sigma(e^+e^- \to \ti{\chi}^0_1 \ti{\chi}^0_3)$
as a function of $\phi_{M_1}$ for scenario A of Table~\ref{scentab},
with $\tan\beta=3$ (solid line),
$\tan\beta=10$ (dashed line), $\tan\beta=30$ (dotted line),
for $\sqrt{s}=500$~GeV and 
transverse beam polarizations 
$({\cal P}_T^-, {\cal P}_T^+)=(100\%,100\%)$. }
\label{fig:plot4}
\end{figure}

Figure~\ref{fig:plot4}a shows the CP asymmetry $A_{CP}$, Eq.~\rf{eq:Asy},
for the process $e^+e^- \to \ti{\chi}^0_1 \ti{\chi}^0_3$ as a
function of $\phi_{M_1}$ for scenario A
defined in Table~\ref{scentab}, for $\tan\beta=3, 10, 30$  
and $\sqrt{s}=500$~GeV. For $\tan\beta=3$~$(30)$ 
the asymmetry $A_{CP}$ reaches its maximum of about $9.6$~$(7.4)\%$ 
at $\phi_{M_1}=1.25$~$(1.55)\pi$. 
Here again the dominant contribution to $A_{CP}$ comes
from the interference term $P(Z \tilde{e}_R)_T$;
see Eq.~\rf{eq:PZeR}.
Note that the maximal CP-violating phase 
$\phi_{M_1}=\frac{\pi}{2}\,(\mbox{mod}\, \pi)$
does not necessarily lead to the highest value of the asymmetry.
The reason for this is an interplay between the $\phi_{M_1}$
dependence of the cross section, shown in Fig. \ref{fig:plot4}b,
and that of the numerator of the asymmetry.
In Fig.~\ref{fig:plot4}b the corresponding cross
section $\sigma(e^+e^- \to \ti{\chi}^0_1 \ti{\chi}^0_3)$
is plotted. For the maximal asymmetry it is about $78$~$(91)$~fb.
In order to measure the asymmetry $A_{CP}$ at 5-$\sigma$,
the required luminosity is  
$\mathcal{L}_{\rm int} \approx 150$~$(217)~\mbox{fb}^{-1}$.

\subsection{Neutralino production and subsequent two-body decays} 

Now we discuss neutralino production $e^+e^- \to \ti{\chi}^0_i \ti{\chi}^0_j$
with the subsequent decays 
$\ti{\chi}^0_j \to \ti{\ell}^\pm_R \ell_1^\mp$ 
and $\ti{\ell}^\pm_R \to \ti{\chi}^0_1 \ell^\pm_2$.
We study the CP-odd asymmetries, 
Eqs.~\rf{eq:AT2}--\rf{eq:ATp}, which are defined by
the azimuthal distribution of the final leptons $\ell_1$ and $\ell_2$.
In this case CP-violation effects arise from the contributions of
the spin correlations of the decaying neutralino,
Eqs.~\rf{eq:SPZeL}--\rf{eq:SPeLeR},
which depend on the transverse beam polarization.
We give numerical examples for $e^+e^- \to \ti{\chi}^0_1 \ti{\chi}^0_2$
and $e^+e^- \to \ti{\chi}^0_1 \ti{\chi}^0_3$.

\subsubsection{CP-odd asymmetries in 
${\bf e^+e^-\to\tilde{\chi}^0_1\tilde{\chi}^0_2}$ production and decay}

\begin{figure}[t]
\setlength{\unitlength}{1mm}
\begin{center}
\begin{picture}(150,35)
\put(-53,-150){\mbox{\epsfig{figure=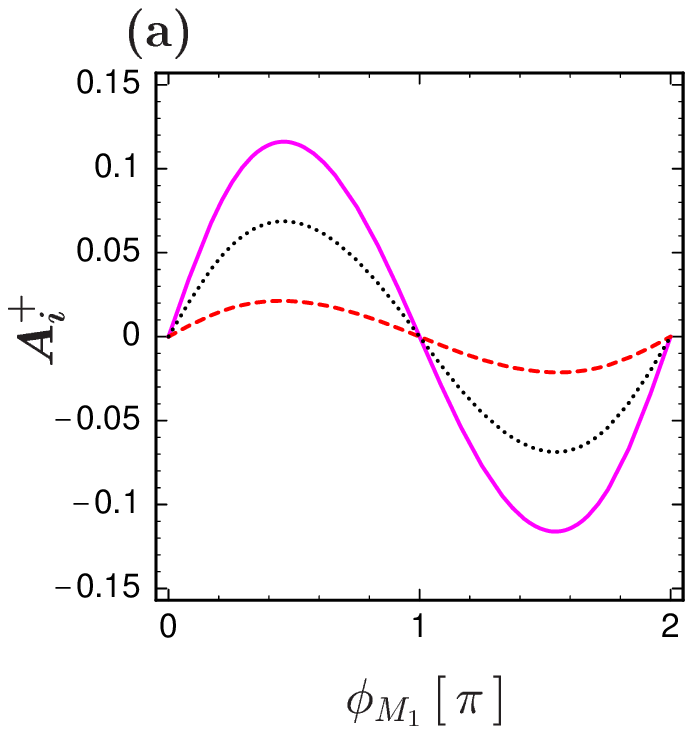,height=22.cm,width=19.4cm}}}
\put(27,-150){\mbox{\epsfig{figure=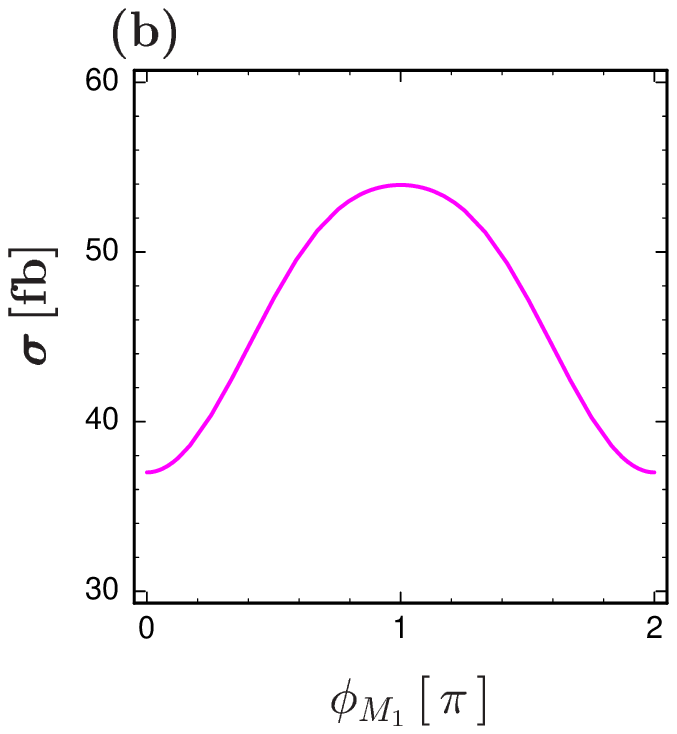,height=22.cm,width=19.4cm}}}
\put(17,36){\small \color{Lila}$A_1^+$}
\put(47,21){\small \color{Red}$A_2^+$}
\put(23,27){\small $A^+$}
\put(19,44){\small ${\bf e^+e^-\to \tilde{\chi}^0_1\tilde{\chi}^0_2
\to \ti{\chi}^0_1 \ti{\chi}^0_1 \ell^\mp_1 \ell^\pm_2}$}
\put(99,44){\small ${\bf e^+e^-\to \tilde{\chi}^0_1\tilde{\chi}^0_2
\to \ti{\chi}^0_1 \ti{\chi}^0_1 \ell^\mp_1 \ell^\pm_2}$}
\end{picture}
\end{center}
\caption{(a) CP-odd asymmetries $A^+_1$ (solid), $A^+_2$ (dashed) and 
$A^+$ (dotted),
Eqs.~\rf{eq:AT2}--\rf{eq:ATp}, for the process 
$e^+e^- \to \ti{\chi}^0_1 \ti{\chi}^0_2
\to \ti{\chi}^0_1 \ti{\chi}^0_1 \ell^\mp_1 \ell^\pm_2$ 
and (b) the corresponding cross section 
as a function of $\phi_{M_1}$ in scenario B (Table~\ref{scentab})
with $\tan\beta=3$.
The centre-of-mass energy is fixed at $\sqrt{s}=500$~GeV and 
the transverse beam polarizations are  
$({\cal P}_T^-, {\cal P}_T^+)=(100\%,100\%)$.
 }
\label{fig:2plot2}
\end{figure}

In Figure~\ref{fig:2plot2}a we show the CP asymmetries $A^+_{1,2}$ and 
$A^+$, Eqs.~\rf{eq:AT2}--\rf{eq:ATp}, as a function of
$\phi_{M_1}$ for scenario B defined in Table~\ref{scentab}.
The beam energy is $\sqrt{s}=500$~GeV with degrees
of transverse beam polarizations $({\cal P}_T^-, {\cal P}_T^+)=(100\%,100\%)$. 
We study neutralino production 
$e^+e^- \to \ti{\chi}^0_1 \ti{\chi}^0_2$, with the subsequent
decays $\ti{\chi}^0_2 \to \ti{\ell}^\pm_R \ell_1^\mp$ 
and $\ti{\ell}^\pm_R \to \ti{\chi}^0_1 \ell^\pm_2$.
For $A^+_1$ we obtain a maximal
value of about $11.6\%$ for $\phi_{M_1}=0.45\pi$.
The asymmetry $A^+_2$ is reduced to $2.1\%$ by the additional
contribution to the phase space from the decay 
$\ti{\ell}^\pm_R \to \ti{\chi}^0_1 \ell^\pm_2$.
Since the branching ratio 
$B(\ti{\ell}^\pm_R \to \ti{\chi}^0_1 \ell^\pm_2)= 1$, we obtain a CP-odd asymmetry 
$A^+\approx 6.9\%$, see Eq.~\rf{eq:ATrel}, for $\phi_{M_1}=0.45\pi$.
In this scenario the main contribution to the CP-odd asymmetries
comes from the $\ti{e}_L$--$\ti{e}_R$ term, Eq.~\rf{eq:SPeLeR}.
With a smaller mass splitting between $m_{\ti{e}_L}$ and $m_{\ti{e}_R}$
the contribution to the asymmetry of $\Sigma^b_P(\ti{e}_L\ti{e}_R)_T$
becomes larger,  but as the cross section is increasing, the 
combination of the two effects leads to a smaller asymmetry.  
The corresponding cross section 
$\sigma(e^+e^- \to \ti{\chi}^0_1 \ti{\chi}^0_2 
\to \ti{\chi}^0_1\ti{\chi}^0_1 \ell_1^\mp\ell_2^\pm)$ 
is plotted
in Fig.~\ref{fig:2plot2}b.  
For $\phi_{M_1}=0.45\pi$, we obtain a cross section 
of about $46$~fb. Thus, for $({\cal P}_T^-, {\cal P}_T^+)=(80\%,60\%)$,
the luminosity $\mathcal{L}_{\rm int}$ needed
for a discovery with 5-$\sigma$ of the asymmetry $A^+_1$ 
is about $176~\mbox{fb}^{-1}$. For a discovery with 5-$\sigma$
of $A^+$, $\mathcal{L}_{\rm int}\approx 517~\mbox{fb}^{-1}$ are needed.

\begin{figure}[t]
\setlength{\unitlength}{1mm}
\begin{center}
\begin{picture}(150,120)
\put(-53,-110){\mbox{\epsfig{figure=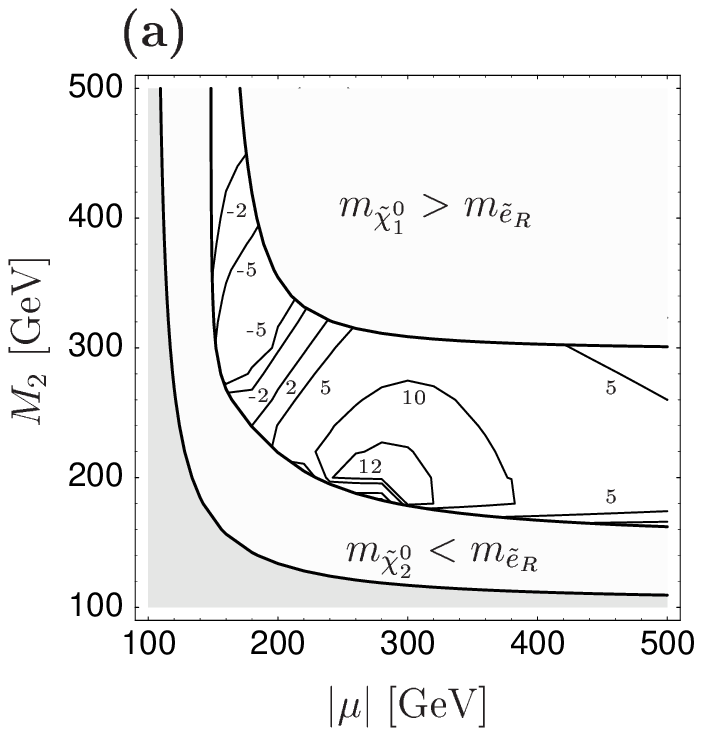,height=27cm,width=19.4cm}}}
\put(27,-110){\mbox{\epsfig{figure=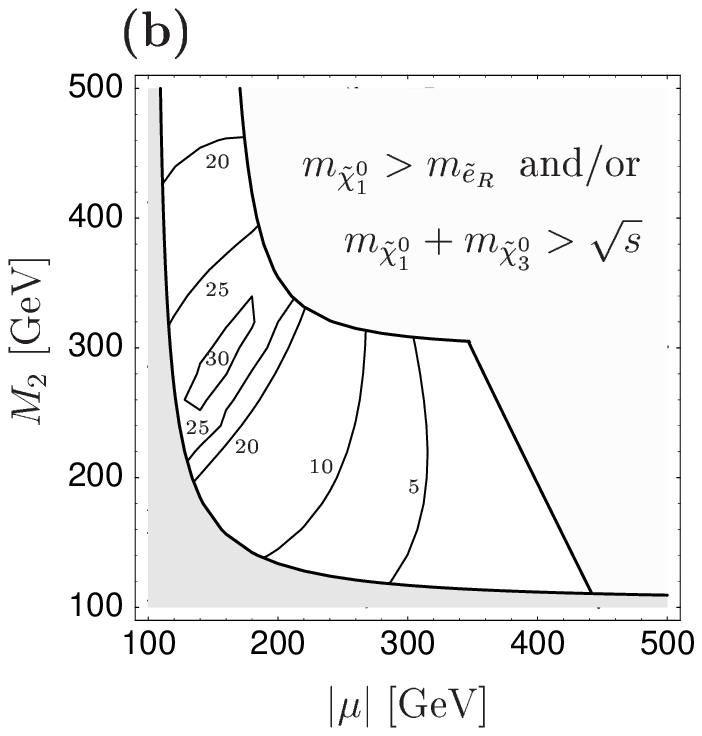,height=27cm,width=19.4cm}}}
\put(27,134){\small ${\bf A^+_1\,[\%]\,\,\mbox{for}}$}
\put(18,129){\small ${\bf e^+e^- \to \ti{\chi}^0_1 \ti{\chi}^0_2
\to \ti{\chi}^0_1 \ti{\ell}^-_R \ell^+_1}$}
\put(107,134){\small ${\bf A^+_1\,[\%]\,\,\mbox{for}}$}
\put(98,129){\small ${\bf e^+e^- \to \ti{\chi}^0_1 \ti{\chi}^0_3
\to \ti{\chi}^0_1 \ti{\ell}^-_R \ell^+_1}$}
\end{picture}
\end{center}
\vspace{-7cm}
\caption{Contours  
of the CP-odd asymmetry $A^+_1$, Eq.~\rf{eq:AT2}, in \% 
in the $|\mu|$--$M_2$ plane
for the process  (a)
$e^+e^- \to \ti{\chi}^0_1 \ti{\chi}^0_2 
\to \ti{\chi}^0_1 \ti{\ell}^-_R \ell^+_1$
and (b)
$e^+e^- \to \ti{\chi}^0_1 \ti{\chi}^0_3 
\to \ti{\chi}^0_1 \ti{\ell}^-_R \ell^+_1$.
The MSSM parameters are 
$\phi_{M_1}=0.5\pi$, $\phi_{\mu}=0$,
$\tan\beta=3$, $m_{\ti{e}_L}=400$~GeV
and $m_{\ti{e}_R}=150$~GeV.
The centre-of-mass energy is fixed at $\sqrt{s}=500$~GeV 
and the transverse beam polarizations are  
$({\cal P}_T^-, {\cal P}_T^+)=(100\%,100\%)$.
The light-grey region is excluded by 
$m_{\ti{\chi}^\pm_1} < 104$~GeV \cite{LEP}.}
\label{fig:2plot1}
\end{figure}

In Fig.~\ref{fig:2plot1}a we show the contour lines 
of the CP-odd asymmetry $A^+_1$, Eq.~\rf{eq:AT2}, for 
$e^+e^- \to \ti{\chi}^0_1 \ti{\chi}^0_2
\to \ti{\chi}^0_1 \ti{\ell}^-_R \ell^+_1$
in the $|\mu|$--$M_2$ plane. The other parameters are 
$\phi_{M_1}=0.5\pi$, $\phi_{\mu}=0$,
$\tan\beta=3$, $m_{\ti{e}_L}=400$~GeV
and $m_{\ti{e}_R}=150$~GeV. The centre-of-mass
energy is fixed at $\sqrt{s}=500$~GeV
with degrees of transverse beam polarizations 
$({\cal P}_T^-, {\cal P}_T^+)=(100\%,100\%)$.
In this figure we only consider the parameter regions where the 
decay channel $\ti{\chi}^0_2 \to \ti{\ell}^\pm_R \ell^\mp_1$
is kinematically accessible.
The maximum value of the CP asymmetry $A^+_1\approx 12.6\%$ is obtained
for a gaugino-like scenario with
$M_2=200$~GeV and $|\mu|=280$~GeV. For these 
parameters the neutralino masses are $m_{\ti{\chi}^0_1}=97$~GeV
and $m_{\ti{\chi}^0_2}=163.5$~GeV, and therefore the branching ratio
$B(\ti{\chi}^0_2 \to \ti{\ell}^\pm_{R}\ell^\mp_1)=1$ and the
cross section
$\sigma(e^+e^- \to \ti{\chi}^0_1 \ti{\chi}^0_2 \to 
\ti{\chi}^0_1 \ti{\ell}^\pm_{R} \ell^\mp_1)=54$~fb. 
Thus for $({\cal P}_T^-, {\cal P}_T^+)=(80\%,60\%)$
the required luminosity $\mathcal{L}_{\rm int}$ for a discovery with 5-$\sigma$, 
Eq.~\rf{eq:estimate}, is about $128~\mbox{fb}^{-1}$.
For this parameter point the asymmetry $A^+$, 
Eq.~\rf{eq:ATp}, is about $6.7\%$ and the luminosity 
$\mathcal{L}_{\rm int}\approx 456~\mbox{fb}^{-1}$.
For higgsino-like scenarios ($M_2>|\mu|$) the $\Sigma^b_P(Z \tilde{e}_R)_{T}$ 
interference term, Eq.~\rf{eq:SPZeR}, gives the main contribution to 
the CP-odd asymmetry, which can be traced back to the structure of the 
corresponding coupling $f^{R*}_{\ell 1}f^{R}_{\ell 2}O^R_{12}$.
On the other hand for gaugino-like scenarios ($|\mu|>M_2$)
the contribution of the interference term 
$\Sigma^b_P(\ti{e}_L \tilde{e}_R)_{T}$ dominates, with
the corresponding coupling
$f^{L*}_{\ell 1}f^L_{\ell 2}f^{R*}_{\ell 1}f^R_{\ell 2}$.
The sign change of the asymmetry in the $|\mu|$--$M_2$ plane
is therefore due to a cancellation of the $Z$--$\ti{e}_R$ and
the $\ti{e}_L$--$\ti{e}_R$ contributions
which have opposite signs.

\subsubsection{CP-odd asymmetries in 
${\bf e^+e^-\to\tilde{\chi}^0_1\tilde{\chi}^0_3}$ production and decay}

In  Fig.~\ref{fig:2plot1}b we show the contour lines of 
the CP asymmetry $A^+_1$ 
for the process $e^+e^- \to \ti{\chi}^0_1 \ti{\chi}^0_3 
\to \ti{\chi}^0_1 \ti{\ell}^-_R \ell^+_1$
in the $|\mu|$--$M_2$ plane.
We fix the MSSM parameters at
$\phi_{M_1}=0.5\pi$, $\phi_{\mu}=0$,
$\tan\beta=3$, $m_{\ti{e}_L}=400$~GeV
and $m_{\ti{e}_R}=150$~GeV with $\sqrt{s}=500$~GeV
and $({\cal P}_T^-, {\cal P}_T^+)=(100\%,100\%)$.  
The maximal CP asymmetry $A^+_1$ is about $31\%$ for $M_2=300$~GeV
and $|\mu|=160$~GeV. For this parameter point we 
obtain neutralino masses of $m_{\ti{\chi}^0_1}=115$~GeV
and $m_{\ti{\chi}^0_2}=156$~GeV, and therefore the branching ratio
is again $B(\ti{\chi}^0_2 \to \ti{\ell}^\pm_{R} \,\ell^\mp_1)=1$.
Hence the cross section
$\sigma(e^+e^- \to \ti{\chi}^0_1 \ti{\chi}^0_3 \to 
\ti{\chi}^0_1 \ti{\ell}^\pm_{R} \,\ell^\mp_1)=83$~fb. 
For transverse beam polarizations 
$({\cal P}_T^-, {\cal P}_T^+)=(80\%,60\%)$
the luminosity $\mathcal{L}_{\rm int}$, Eq.~\rf{eq:estimate}, 
for a discovery with 5-$\sigma$ of $A^+_1$ is about $14~\mbox{fb}^{-1}$.
For these parameters the CP-odd asymmetry $A^+$ is
$17.3\%$ and the necessary luminosity 
(for a discovery with 5-$\sigma$) $\mathcal{L}_{\rm int}\approx 43~\mbox{fb}^{-1}$.
In the case of $e^+e^- \to \ti{\chi}^0_1 \ti{\chi}^0_3$
the CP-violating contributions  $\Sigma^b_P(Z \tilde{e}_R)_{T}$ and
$\Sigma^b_P(\tilde{e}_L \tilde{e}_R)_{T}$, 
Eqs.~\rf{eq:SPZeR} and \rf{eq:SPeLeR}, enter with the same sign due
to the corresponding couplings. In gaugino-like scenarios the
contributions of both interference terms are suppressed,
therefore the asymmetry decreases.

\begin{figure}[t]
\setlength{\unitlength}{1mm}
\begin{center}
\begin{picture}(150,35)
\put(-53,-150){\mbox{\epsfig{figure=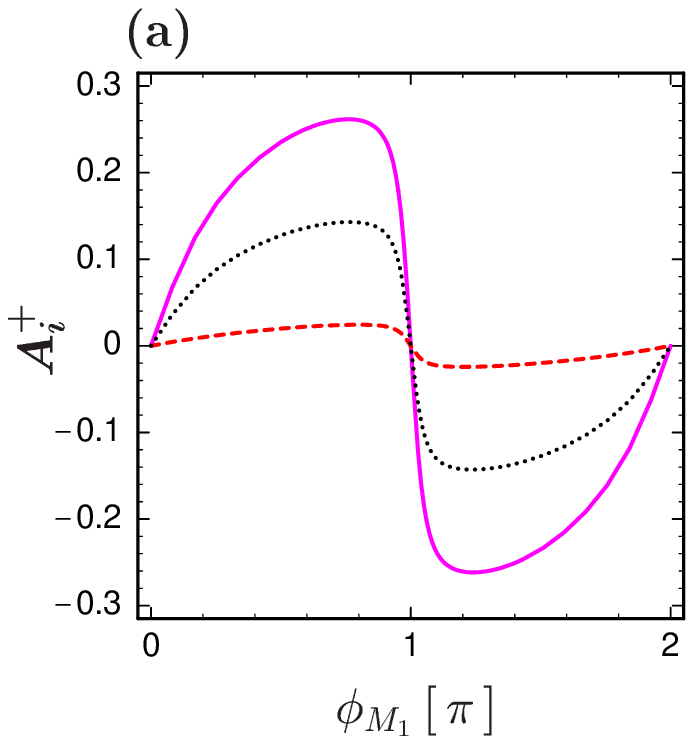,height=22.cm,width=19.4cm}}}
\put(27,-150){\mbox{\epsfig{figure=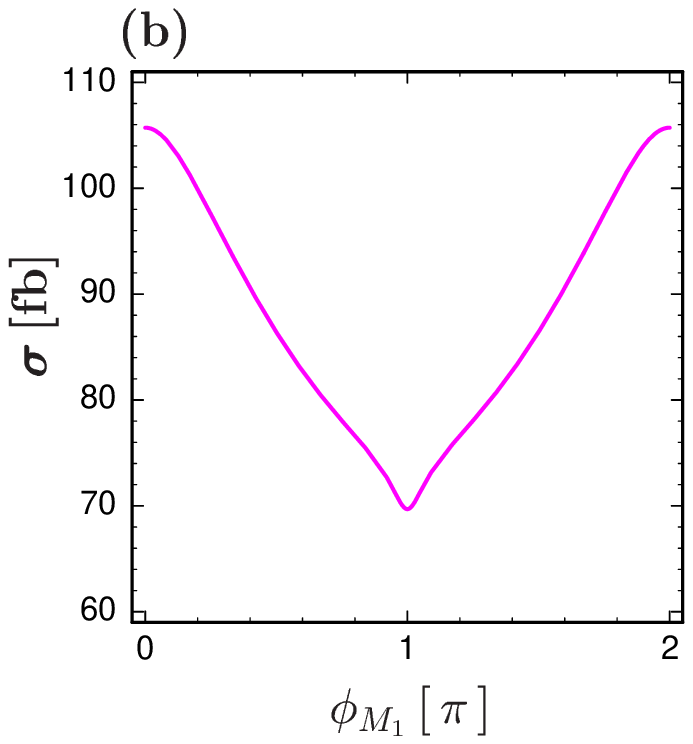,height=22.cm,width=19.4cm}}}
\put(17,37){\small \color{Lila}$A_1^+$}
\put(47,23){\small \color{Red}$A_2^+$}
\put(28,27){\small $A^+$}
\put(19,44.5){\small ${\bf e^+e^-\to \tilde{\chi}^0_1\tilde{\chi}^0_3
\to \ti{\chi}^0_1 \ti{\chi}^0_1 \ell^\mp_1 \ell^\pm_2}$}
\put(99,44.5){\small ${\bf e^+e^-\to \tilde{\chi}^0_1\tilde{\chi}^0_3
\to \ti{\chi}^0_1 \ti{\chi}^0_1 \ell^\mp_1 \ell^\pm_2}$}
\end{picture}
\end{center}
\caption{CP-odd asymmetries $A^+_1$ (solid), $A^+_2$ (dashed)
and $A^+$ (dotted), 
Eqs.~\rf{eq:AT2}--\rf{eq:ATp},
for the process 
$e^+e^- \to \ti{\chi}^0_1 \ti{\chi}^0_3
\to \ti{\chi}^0_1 \ti{\chi}^0_1 {\ell}^\mp_1 \ell^\pm_2$
and (b) the corresponding cross section as a function of $\phi_{M_1}$
in scenario A, see Table~\ref{scentab}, with $\tan\beta=3$.
The centre-of-mass energy is $\sqrt{s}=500$~GeV 
and the transverse beam polarizations are 
$({\cal P}_T^-, {\cal P}_T^+)=(100\%,100\%)$.
 }
\label{fig:2plot3}
\end{figure}

In Fig.~\ref{fig:2plot3}a the CP-odd asymmetries $A^+_{1,2}$ and $A^+$, 
Eqs.~\rf{eq:AT2}--\rf{eq:ATp}, for the process 
$e^+e^- \to \ti{\chi}^0_1 \ti{\chi}^0_3
\to\ti{\chi}^0_1\ti{\chi}^0_1 \ell^\mp_1 \ell^\pm_2$ are
plotted as a function of $\phi_{M_1}$ for scenario A, 
see Table~\ref{scentab}. 
For a centre-of-mass energy  $\sqrt{s}=500$~GeV and with
transverse beam polarization $({\cal P}_T^-, {\cal P}_T^+)=(100\%,100\%)$
the maximum of the CP asymmetry $A^+_1(A^+)\approx 26.2$~$(14.3)\%$ is obtained
for $\phi_{M_1}\approx 0.75\pi$. 
In this scenario we have large mixing between the gaugino
and the higgsino components, the main contribution
to $A^+_1$ stems again from the
$Z$--$\ti{e}_R$ interference term, which is about $21.4\%$. 
The $\ti{e}_R$--$\ti{e}_L$ contribution is $4\%$, wheras
the $Z$--$\ti{e}_L$ contribution is suppressed by the large mass of
the left selectron.
Figure~\ref{fig:2plot3}b 
shows the cross section
$\sigma(e^+e^- \to \ti{\chi}^0_1 \ti{\chi}^0_3
\to \ti{\chi}^0_1 \ti{\ell}^\pm_R \ell^\mp_1)$.
For the maximum of the CP asymmetries $A^+_1(A^+)$, for 
$\phi_{M_1}\approx 0.75\pi$, the cross section is $78$~fb.
For $({\cal P}_T^-, {\cal P}_T^+)=(80\%,60\%)$ 
the luminosity $\mathcal{L}_{\rm int}$ for a discovery with 5-$\sigma$ 
is about $20$~$(68)~\mbox{fb}^{-1}$.

\subsection{Determination of the SUSY parameters} 

In the following we will give an example for the accuracy
that can be expected in the determination of the MSSM
parameters, focusing on the determination of 
the complex parameter $M_1=|M_1|e^{i\,\phi_{M_1}}$.
In order to determine
the parameters unambiguously, CP-even as well as
CP-odd observables have to be included in the set of observables
from which the underlying parameters are extracted.
\vspace{-0.5cm}
\begin{table}[H]
\begin{center}
\begin{tabular}{|c||c|c|c|c|c|c|c|c|} \hline
$({\cal P}^-_L,{\cal P}^+_L)$
& \multicolumn{1}{|c|}{$\qquad(0,0)\qquad$} 
& \multicolumn{1}{c|}{$(-80\%,+60\%)$} 
& \multicolumn{1}{c|}{$(+80\%,-60\%)$}\\\hline\hline
$\sigma(e^+e^- \to \ti{\chi}^0_1 \ti{\chi}^0_2)$
& \multicolumn{1}{|c|}{47.27~fb} 
& \multicolumn{1}{c|}{87.13~fb} 
& \multicolumn{1}{c|}{52.80~fb}\\\hline
$\sigma(e^+e^- \to \ti{\chi}^0_2 \ti{\chi}^0_2)$
&\multicolumn{1}{|c|}{11.59~fb} 
& \multicolumn{1}{c|}{33.12~fb}  
& \multicolumn{1}{c|}{1.186~fb}\\\hline
$\sigma(e^+e^- \to \ti{\chi}^0_1 \ti{\chi}^0_3)$
&\multicolumn{1}{|c|}{9.83~fb} 
& \multicolumn{1}{c|}{5.68~fb}  
& \multicolumn{1}{c|}{23.42~fb}\\\hline
$\sigma(e^+e^- \to \ti{\chi}^0_1 \ti{\chi}^0_4)$
&\multicolumn{1}{|c|}{7.86~fb} 
& \multicolumn{1}{c|}{7.74~fb}  
& \multicolumn{1}{c|}{15.53~fb}\\\hline
\end{tabular}\\[1.5ex]
\caption{\label{tab3}
Cross sections for different sets of longitudinal
beam polarizations in scenario B with 
$\phi_{M_1}=0.5\pi$ and $\phi_\mu=0$ for 
$\sqrt{s}=500$~GeV.}
\end{center}
\end{table}
\vspace{-0.5cm}
Our set of observables contains the neutralino masses 
$m_{\tilde{\chi}^0_j}$, the cross sections
$e^+e^-\to\ti\chi^0_i\ti\chi^0_j$ for different
choices of longitudinal beam polarizations
$({\cal P}^-_L, {\cal P}^+_L)=(0,0)$, $(-80\%,+60\%)$, $(+80\%,-60\%)$
and the CP-odd asymmetry $A_{CP}$, Eq.~\rf{eq:Asy}.
We now take scenario B with
$\phi_{M_1}=0.5\pi$ and $\phi_\mu=0$, see Table~\ref{scentab}, as our
reference point of input parameters.
We calculate the neutralino masses
$m_{\tilde{\chi}^0_1}$, $m_{\tilde{\chi}^0_2}$,
$m_{\tilde{\chi}^0_3}$ and $m_{\tilde{\chi}^0_4}$, see Table~\ref{scentab}. 
The cross sections 
for $e^+e^-\to\ti\chi^0_1\ti\chi^0_2$, $e^+e^-\to\ti\chi^0_2\ti\chi^0_2$,
$e^+e^-\to\ti\chi^0_1\ti\chi^0_3$ and $e^+e^-\to\ti\chi^0_1\ti\chi^0_4$
for $\sqrt{s}=500$~GeV with different sets of longitudinal 
beam polarizations are displayed in Table~\ref{tab3}. 
The CP asymmetry $A_{CP}$ for the process 
$e^+e^-\to\ti\chi^0_1\ti\chi^0_2$ is about $+2\%$ for
transverse beam polarizations $({\cal P}_T^-, {\cal P}_T^+)=(80\%,60\%)$.
We regard these
calculated values as real experimental data, where we assume
errors of $1\%$ for the masses. For the error of the cross sections
of each polarization configuration 
and of the asymmetry we take a 1-$\sigma$ deviation for
a luminosity $\mathcal{L}_{\rm int}=100~\mbox{fb}^{-1}$.
Our approach for the determination of the error of the parameters
(in particular the error of $M_1$) is described as follows:
we perform a random scan over the input parameters $|M_1|$, $\phi_{M_1}$,
$M_2$, $|\,\mu\,|$, $\phi_\mu$ and $\tan\beta$ around
our reference point and select the points which pass the
condition $|\mathcal{O}^{\rm meas}_i-\mathcal{O}^{\rm calc}_i|
<|\Delta \mathcal{O}^{\rm meas}_i|$,
where $\mathcal{O}^{\rm meas}_i$ are the values of the
observables at our reference point, see Table~\ref{tab3},
$\Delta \mathcal{O}^{\rm meas}_i$ is the corresponding
error, and $\mathcal{O}^{\rm calc}_i$ are the values of the calculated 
observables obtained through the random scan.
\begin{figure}[t]
\setlength{\unitlength}{1mm}
\begin{center}
\begin{picture}(150,120)
\put(-15,-110){\mbox{\epsfig{figure=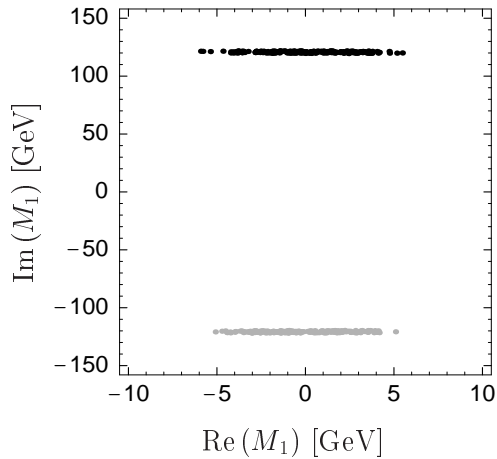,height=27cm,width=19.4cm}}}
\end{picture}
\end{center}
\vspace{-7cm}
\caption{SUSY parameter points in the $\mbox{Re}(M_1)$--$\mbox{Im}(M_1)$ 
plane consistent with scenario B, if one assumes an uncertainty of $1\%$ 
for the masses and 1-$\sigma$ deviation 
($\mathcal{L}_{\rm int}=100~\mbox{fb}^{-1}$)
for the cross sections and the asymmetry. 
The parameters $|M_1|$, $\phi_{M_1}$,
$M_2$, $|\,\mu\,|$, $\phi_\mu$ and $\tan\beta$ have
been randomly scanned around the reference point B.  
The grey points
are excluded if one takes into account the CP-odd observable $A_{CP}$,
Eq.~\rf{eq:Asy}.}
\label{fig:fehler}
\end{figure}  

In Fig.~\ref{fig:fehler} we show the SUSY parameter points 
compatible with our reference scenario in
the $\mbox{Re}(M_1)$--$\mbox{Im}(M_1)$ plane.  
If we consider only the CP-even
observables (cross sections and masses) then we obtain two regions
in the parameter space compatible with our reference scenario. 
This ambiguity can be resolved if one includes
in addition the CP-odd observable $A_{CP}$, Eq.~\rf{eq:Asy}.
For the error of $M_1$ we obtain: $\mbox{Re}(M_1)=0 \pm 5.9$~GeV
and $\mbox{Im}(M_1)=120.8 \pm 1.3$~GeV.

\section{Conclusion \label{conclusion}}

We have studied the processes $e^+e^-\to\ti\chi^0_1\ti\chi^0_2$
and $e^+e^-\to\ti\chi^0_1\ti\chi^0_3$ with subsequent decays
$\ti{\chi}^0_{2,3}\to\ti{\ell}_{R}\ell$ and 
$\ti{\ell}_{R}\to\ti{\chi}^0_1\ell$, where $\ell=e,\mu$,
at a linear collider with transverse $e^+$ and $e^-$ beam polarizations.
We have discussed different CP asymmetries, which are due to azimuthal
distributions of the neutralinos or the final leptons.
We have pointed out that these CP asymmetries are non-vanishing
thanks to the Majorana character of the neutralinos.
We have given the analytical expressions for the CP asymmetries 
and the cross sections in the spin density matrix formalism, including
the complete spin correlations between production and decay of
the neutralinos. At the ILC
at $\sqrt{s}=500$~GeV and with degrees of transverse
$e^\pm$ beam polarizations
$({\cal P}^-_T,{\cal P}^+_T)=(80\%,60\%)$, the CP asymmetries 
can reach up to about $15\%$. Also we have shown 
that these CP asymmetries can be observed in a broad 
range of the MSSM parameter space. Furthermore, we have discussed
the unambiguous determination of the underlying SUSY parameters, which
requires CP-even as well as CP-odd observables.

\section*{Acknowledgements}
A.$\,$B. is grateful to S.$\,$Rindani for valuable discussions.
Furthermore we thank O.$\,$Kittel and A.$\,$Wagner for interesting
discussions. S.$\,$H. acknowledges the hospitality of the IPPP, Durham.
This work is supported by the `Fonds zur
F\"orderung der wissenschaftlichen Forschung' (FWF) of Austria, project
No. P16592-N02, by the Deutsche Forschungsgemeinschaft (DFG) under
contract No.\ \mbox{FR~1064/5-2}, and by the European Community's 
Human Potential Programme under contract HPRN--CT--2000--00149.  
S.H.\ is supported by the G\"oran Gustafsson Foundation.

\section*{Appendices}
\begin{appendix}
\section{Momentum and polarization vectors\label{appendixA}}

The basis vectors for transverse $e^-$ beam polarization are 
\begin{eqnarray}
\vec{t}_{e^-}^{\,\, 1}&=&(\vec{t}^{\,\, 2}_{e^-}\times \vec{p}_{e^-})/|\vec{t}^{\,\, 2}_{e^-}\times \vec{p}_{e^-}|~,\\
\vec{t}^{\,\, 2}_{e^-}&=&(\vec{p}_{e^-}\times \vec{p}_{\chi_j})/|\vec{p}_{e^-}\times \vec{p}_{\chi_j}|~.
\end{eqnarray}
The basis vectors for transverse $e^+$ 
beam polarization are defined analogously.
In a fixed coordinate system $(x,y,z)$, with
the $z$-axis pointing along the beam direction 
the basis vectors in the cms are given by
\begin{eqnarray} 
t^1_{e^\pm}&=&(0,\cos\phi,\sin\phi,0)\quad\mbox{and}\quad t^2_{e^\pm}=(0,-\sin\phi,\cos\phi,0)~. 
\label{eq_tem} 
\end{eqnarray}
The momentum 4-vectors of $\ti\chi^0_i$ and $\ti{\chi}^0_j$ are 
\begin{eqnarray}
\label{eq:momentumneut}
p_{\chi_j,\,\mu}&=&p_{4,\,\mu}=q
(E_{\chi_j}/q,\cos\phi \sin\theta,\sin\phi \sin\theta,\cos\theta)~,\nonumber\\
p_{\chi_i,\,\mu}&=&p_{3,\,\mu}=q
(E_{\chi_i}/q,-\cos\phi \sin\theta,-\sin\phi \sin\theta,-\cos\theta)~,
\end{eqnarray}
with
\be{eq:energy}
E_{\chi_{i,j}}=\frac{s+m^2_{\chi_{i,j}}-m^2_{\chi_{j,i}}}{2 \sqrt{s}}~,\qquad
q=\frac{\lambda^{\frac{1}{2}}(s,m^2_{\chi_i},m^2_{\chi_j})}{2 \sqrt{s}}~,
\ee
where $\lambda(a,b,c)=a^2+b^2+c^2-2(a b + a c + b c)$.
The three spin-basis vectors $s^b_{\chi_j,\,\mu}$ of $\ti\chi^0_j$
are chosen to be 
\baq{eq:polvec}
s^1_{\chi_j,\,\mu}&=&\left(0,\frac{{\vec{s}}_2\times{\vec{s}}_3}
{|{\vec{s}}_2\times{\vec{s}}_3|}\right)=
(0,-\cos\phi \cos\theta,-\sin\phi \cos\theta,\sin\theta)~,
\nonumber \\[3mm]
s^2_{\chi_j,\,\mu}&=&\left(0,
\frac{{\vec{p}}_{\chi_j}\times{\vec{p}}_{e^-}}
{|{\vec{p}}_{\chi_j}\times{\vec{p}}_{e^-}|}\right)=
(0,\sin\phi,-\cos\phi,0)~,
\nonumber \\[3mm]
s^3_{\chi_j,\,\mu}&=&\frac{1}{m_{\chi_j}}
\left(q, 
\frac{E_{\chi_j}}{q}{\vec{p}}_{\chi_j} \right)=
\frac{E_{\chi_j}}{m_{\chi_j}}
(q/E_{\chi_j},\cos\phi \sin\theta,\sin\phi \sin\theta,\cos\theta)~,
\eaq
where $\vec{s}^{\,\,1}_{\tilde{\chi}_j}$, $\vec{s}^{\,\,2}_{\tilde{\chi}_j}$ 
and $\vec{s}^{\,\,3}_{\tilde{\chi}_j}$ build a right-handed-system.
The momentum 4-vector of the lepton in the decay $\ti\chi^0_j
\to\tilde{\ell}_{L,R}\, \ell$ is given by
\be{eq:fourlep}
p_{\ell_1,\,\mu}=
|{\vec{p}_{\ell_1}}| (1,\cos\phi_{\ell_1} \sin\theta_{\ell_1},
\sin\phi_{\ell_1} \sin\theta_{\ell_1},
\cos\theta_{\ell_1})
\ee
with
\be{eq:lepmom}
|{\vec{p}_{\ell_1}}|=\frac{m^2_{\chi_j}-
m^2_{\ti{\ell}}}{2(E_{\chi_j}-q\cos\vartheta)}
\ee
and
\be{eq:angle}
\cos\vartheta=\sin\theta \sin\theta_{\ell_1} \cos(\phi-\phi_{\ell_1})+
\cos\theta \cos\theta_{\ell_1}~.
\ee
The momentum 4-vector of the lepton from the 
decay $\ti\ell\to\ti\chi^0_1\ell_2$ is given by
\be{eq:fourlep2}
p_{\ell_2,\,\mu}=
|{\vec{p}_{\ell_2}}| (1,\cos\phi_{\ell_2} \sin\theta_{\ell_2},
\sin\phi_{\ell_2} \sin\theta_{\ell_2},
\cos\theta_{\ell_2})~,
\ee
where
\be{eq:lepmom2}
|{\vec{p}_{\ell_2}}|=\frac{m^2_{\ti\ell}-
m^2_{\chi_1}}{2(E_{\ti\ell}-|{\vec{p}_{\ti\ell}}|~
({\hat{\vec{p}}_{\ti\ell}\cdot}{\hat{\vec{p}}_{\ell_2}}))}~.
\ee

\section{Decay matrix and phase space of 2-body decay \label{appendixB}}

The spin density matrix
of the decay $\ti{\chi}^0_j \to \ti{\ell}^\mp_{L,R} \,\ell^\pm$
can be written as
\be{eq:decdensity}
\rho_{D,\lambda'_j\lambda_j}=\delta_{\lambda'_j\lambda_j}D(\ti{\chi}^0_j)+
\sum^3_{c=1} \sigma^c_{\lambda'_j\lambda_j} \Sigma^c_D(\ti{\chi}^0_j)~,
\ee
where the expansion coefficient $D(\ti \chi^0_j)$ 
is the part that is independent of the 
polarization of the decaying neutralino $\ti{\chi}^0_j$, and 
$\Sigma^a_D(\ti \chi^0_j)$ is the part that depends 
on the polarization of $\ti{\chi}^0_j$.
For the sake of simplicity we consider $\ell=e,\mu$, where the mixing
in the slepton sector can be neglected. 
Then we have, for 
$\ti \chi^0_j \to \ti \ell^\mp_{L}~\ell^\pm$:  
\be{eq:DslepL}
D(\ti \chi^0_j \to \ti\ell^\mp_L~\ell^\pm)= \frac{g^2}{2} |f^L_{\ell j}|^2 
(m^2_{\chi_j}-m^2_{\ti{\ell}_L})~,
\ee
\be{eq:SDslepL}
\Sigma^c_D(\ti \chi^0_j \to \ti\ell^\mp_L~\ell^\pm)=\mp g^2 |f^L_{\ell j}|^2 
m_{\chi_j} (s^c \cdot p_{\ell})
\ee
and for $\ti \chi^0_j \to \ti \ell^\mp_R~\ell^\pm$  
\be{eq:DslepR}
D(\ti \chi^0_j \to \ti\ell^\mp_R~\ell^\pm)=\frac{g^2}{2} |f^R_{\ell j}|^2 
(m^2_{\chi_j}-m^2_{\ti{\ell}_R})~,
\ee
\be{eq:SDslepR}
\Sigma^c_D(\ti \chi^0_j \to \ti\ell^\mp_R~\ell^\pm)=\pm g^2 |f^R_{\ell j}|^2 
m_{\chi_j} (s^c \cdot p_{\ell})~,
\ee
where $m_{\chi_j}$ ($m_{\ti{\ell}_{L,R}}$) is the mass of
$\ti{\chi}^0_j$ ($\ti{\ell}_{L,R}$).
The parametrizations of the momentum
4-vector $p_{\ell,\,\mu}$ and the 
polarization 4-vector $s^c_{\ti{\chi}_j,\,\mu}$
of the neutralino $\ti{\chi}^0_j$ are given in 
Eqs.~(\ref{eq:momentumneut}) and \rf{eq:polvec} in 
Appendix \ref{appendixA}.
Finally, the matrix element squared for the two-body decay 
$\ti\ell^\mp \to \ti\chi^0_1~ \ell^\mp$ 
in the decay chain, Eq.~(\ref{eq:decaychain}), is 
\be{eq:D2}
D_2(\ti\ell^\mp_{L,R}\to \ti\chi^0_1~\ell^\pm)= g^2 |f^{L,R}_{\ell 1}|^2~
(m^2_{\ti{\ell}_{L,R}}-m^2_{\chi_1})~.
\ee
From (\ref{eq4_5}) and \rf{eq:decdensity},
summing over the polarization
of $\ti\chi^0_i$, whose decay is not considered, the 
differential cross section for  
$e^+e^-\to \ti\chi^0_i\ti\chi^0_j\to
\ti\chi^0_i\ti\ell^\mp_{L,R}~\ell^\pm_1$ is:
\be{eq:crossection}
{\rm d}\sigma_1=\frac{2}{s}\left[P D+\Sigma^a_P \Sigma^a_D\right]~
|\Delta(\ti\chi^0_j)|^2 {\rm dLips}_1~.
\ee
Similarly one obtains the differential cross section for 
$e^+e^-\to \ti\chi^0_i\ti\chi^0_j\to
\ti\chi^0_i\ti\ell^\mp_{L,R}~\ell^\pm_1 \to
\ell^\pm_1\ell^\mp_2\ti\chi^0_1\ti\chi^0_i$ from
(\ref{eq4_5}), \rf{eq:decdensity} and \rf{eq:D2}
and summing over the polarization
of $\ti\chi^0_i$:
\be{eq:crossection2}
{\rm d}\sigma_2=\frac{2}{s}\left[P D+\Sigma^a_P \Sigma^a_D\right] D_2~
|\Delta(\ti\chi^0_j)|^2 |\Delta(\ti\ell)|^2 {\rm dLips}_2~,
\ee
where $P$ and $\Sigma^a_P$ involve the terms for
arbitrary beam polarization.
For the calculation of the cross section we use
the narrow widths approximation 
($\int |\Delta(\ti\chi^0_j)|^2 {\rm d}\hat{s}_{\ti{\chi}_j}
=\frac{\pi}{m_{\ti{\chi}_j}\Gamma_{\ti{\chi}_j}}$
and 
$\int |\Delta(\ti\ell)|^2 {\rm d}\hat{s}_{\ti\ell}
=\frac{\pi}{m_{\ti\ell}\Gamma_{\ti\ell}}$, where
${\rm d}\hat{s}_{\ti{\chi}_j}=p^2_{\ti{\chi}_j}$
and
${\rm d}\hat{s}_{\ti{\ell}}=p^2_{\ti{\ell}}$). 
The Lorentz-invariant phase-space elements
in Eqs.~\rf{eq:crossection} and \rf{eq:crossection2} 
for the decay chain
$\ti\chi^0_j\to\ti\ell^\mp_{L,R} \ell^\pm_1
\to \ti\chi_1^0 \ell^\pm_1 \ell^\mp_2$ 
are
\be{eq:phasespace1}
{\rm d Lips}_1 =\frac{1}{2 \pi} {\rm d Lips}(s,p_{\chi_i},p_{\chi_j})
{\rm d}\hat{s}_{m_{\chi_j}}
{\rm d Lips}(\hat{s}_{m_{\chi_j}},p_{\ti\ell},p_{\ell_1})~,
\ee
\be{eq:phasespace2}
{\rm d Lips}_2 =\frac{1}{(2 \pi)^2} {\rm d Lips}(s,p_{\chi_i},p_{\chi_j})
{\rm d}\hat{s}_{m_{\chi_j}}
{\rm d Lips}(\hat{s}_{m_{\chi_j}},p_{\ti\ell},p_{\ell_1})
{\rm d}\hat{s}_{m_{\ti\ell}}
{\rm d Lips}(\hat{s}_{m_{\ti\ell}},p_{\chi_1},p_{\ell_2})
\ee
with the Lorentz invariant phase space elements 
\be{eq:prodphs}
{\rm d Lips}(s,p_{\chi_i},p_{\chi_j})=\frac{1}{4(2 \pi)^2} \frac{q}{\sqrt{s}}
\sin\theta~ {\rm d}\theta~ {\rm d}\phi~,
\ee
\be{eq:lepphs}
{\rm d Lips}(\hat{s}_{m_{\chi_j}},p_{\ti\ell},p_{\ell_1})=
\frac{1}{2(2 \pi)^2}\frac{|{\vec{p}}_{\ell_1}|}{m^2_{\chi_j}-m^2_{\ti\ell}}
\sin\theta_{\ell_1}~ {\rm d}\theta_{\ell_1}~ {\rm d}\phi_{\ell_1}~,
\ee
\be{eq:lepphs2}
{\rm d Lips}(\hat{s}_{m_{\ti\ell}},p_{\chi_1},p_{\ell_2})=
\frac{1}{2(2 \pi)^2}\frac{|{\vec{p}}_{\ell_2}|}{m^2_{\ti\ell}-m^2_{\chi_1}}
\sin\theta_{\ell_2}~ {\rm d}\theta_{\ell_2}~ {\rm d}\phi_{\ell_2}~.
\ee

\section{Reconstruction of the production plane \label{appendixC}}

As an example, we consider the process $e^+e^-\to\ti\chi^0_1\ti\chi^0_2$
with the decays 
$\ti\chi^0_2\to \ell_1~ \ti\ell$ and $\ti\ell \to \ell_2~ \ti\chi'^{\,0}_1$,
where we denote the neutralino from the decay by $\ti\chi'^{\,0}_1$
(here again the labels of the leptons indicate their origin).
We assume that the masses of all particles involved are known.

We rotate to a coordinate system where the $3$-momentum vector of $\ell_1$
is along the $z$-axis and that of $\ell_2$ is
in the $x$--$z$ plane. The unit vectors of the $3$-momenta 
of $\ell_1$, $\ell_2$, $\ti\ell$ are 
\be{eq:unitvec}
\hat{\vec{p}}_{\ell_1}=(0,0,1)~,\quad \hat {\vec{p}}_{\ell_2}=(\sin c,0,\cos c)~,
\quad \hat {\vec{p}}_{\ti\ell}=(\sin b \cos A,\sin b \sin A,\cos b).
\ee
From the relation 
$({\vec{p}}_{\ell_1}+{\vec{p}}_{\ti\ell})^2={\vec{p}}_{\chi_2}^{\,\,2}$
we obtain
\be{eq:cosb}
\cos b=\frac{1}{2|{\vec{p}}_{\ell_1}||{\vec{p}}_{\ti\ell}|}
\left[
|{\vec{p}}_{\chi_2}|^2-|{\vec{p}}_{\ell_1}|^2-|{\vec{p}}_{\ti\ell}|^2
\right]~,
\ee
where $|{\vec{p}}_{\ti\ell}|^2=E^2_{\ti\ell}-m^2_{\ti\ell}$ and
$E_{\ti\ell}=E_{\chi_2}-E_{\ell_1}$.
From a second relation 
$({\vec{p}}_{\ti\ell}-{\vec{p}}_{\ell_2})^2={\vec{p}}_{\chi'_1}^{\,\,2}$ 
between momentum vectors we obtain
\be{eq:cosa}
\cos a=\frac{1}{2|{\vec{p}}_{\ell_2}||{\vec{p}}_{\ti\ell}|}
\left[
|{\vec{p}}_{\ti\ell}|^2+|{\vec{p}}_{\ell_2}|^2-|{\vec{p}}_{\chi'_1}|^2
\right]~,
\ee
where $|{\vec{p}}_{\chi'_1}|^2=E^2_{\chi'_1}-m^2_{\chi_1}$ and
$E_{\chi'_1}=E_{\chi_2}-E_{\ell_1}-E_{\ell_2}$ or 
$E_{\chi'_1}=\not\!\! E-E_{\chi_1}$, and $\not\!\! E$
is the missing energy.
From spherical geometry with
$\hat{\vec{p}}_{\ell_2}\cdot\hat{\vec{p}}_{\ti\ell}=\cos a$, we
obtain the following relation 
between the angles
\be{eq:relangles}
\cos A=\frac{\cos a-\cos b~ \cos c}{\sin c~ \sin b}~.
\ee
Inserting \rf{eq:cosb}, \rf{eq:cosa} and \rf{eq:relangles}
into $\hat{\vec{p}}_{\ti\ell}$ in \rf{eq:unitvec}, this vector
is determined up to a twofold ambiguity in the second component.
In order to resolve this ambiguity, a reference vector is needed,
which tells us in which hemisphere of the $x$--$z$ plane the momentum
vector $\hat{\vec{p}}_{\ti\ell}$ is. For instance, 
this is possible in the process 
$e^+e^-\to\ti\chi^0_1\ti\chi^0_3$ with the decays
$\ti\chi^0_3\to \ell_1~ \ti\ell$,
$\ti\ell \to \ell_2~ \ti\chi^{\,0}_2$ and
$\ti\chi^0_2\to \ti\chi'^{\,0}_1 Z$, where the $3$-momentum
of the Z boson is the reference vector.

\end{appendix}

\end{document}